\documentclass[aps,prb,twocolumn,longbibliography]{revtex4-2}  

\usepackage{amsmath,amssymb,amsfonts,bm,physics,latexsym,comment} 

\usepackage[pdftex]{graphicx,color}
\usepackage{hyperref}
\hypersetup{
   colorlinks=true,        
   linkcolor=blue,          
   citecolor=blue,        
  filecolor=magenta,      
   urlcolor=blue           
}
\usepackage{tikz}

\newcommand{\Jperp}{J_\perp}
\newcommand{\Jdiag}{J_\times}
\newcommand{\Jpd}{J_{\perp,\times}}

\newcommand{\Sv}{\bm{S}}
\newcommand{\Tv}{\bm{T}}

\newcommand{\Ucal}{{\cal U}}

\newcommand{\ua}{\uparrow}
\newcommand{\da}{\downarrow}

\begin{document}
\title{
Symmetry, topology, duality, chirality, and criticality\\ in a spin-$\frac12$ XXZ ladder with a four-spin interaction
}
\author{Mat\'eo Fontaine}
\affiliation{Department of Physics, Keio University, Kohoku-ku, Yokohama, Kanagawa 223-8522, Japan}
\author{Koudai Sugimoto}
\affiliation{Department of Physics, Keio University, Kohoku-ku, Yokohama, Kanagawa 223-8522, Japan}
\author{Shunsuke Furukawa}
\affiliation{Department of Physics, Keio University, Kohoku-ku, Yokohama, Kanagawa 223-8522, Japan}

\date{\today}

\begin{abstract}
We study the ground-state phase diagram of a spin-$\frac12$ XXZ model with a chirality-chirality interaction (CCI) on a two-leg ladder. 
This model offers a minimal setup to study an interplay between spin and chirality degrees of freedom. 
The spin-chirality duality transformation allows us to relate the regimes of weak and strong CCIs. 
By applying the Abelian bosonization and the duality, 
we obtain a rich phase diagram that contains distinct gapped featureless and ordered phases. 
In particular, N\'eel and vector chiral orders appear for easy-axis anisotropy, 
while two distinct symmetry protected topological (SPT) phases appear for easy-plane anisotropy. 
The two SPT phases can be viewed as twisted variants of the Haldane phase. 
We also present an effective description in terms of (spinor) hard-core bosons, 
which reveals critical behavior on the self-dual line in the easy-axis and easy-plane regimes. 
We perform numerical simulations to confirm the predicted phase structure and critical properties. 
We further demonstrate that the two SPT phases and a trivial phase are distinguished by topological indices in the presence of certain symmetries. 
A similar phase structure is expected in a spin-$\frac12$ XXZ ladder with four-spin ring exchange.  
\end{abstract}
\maketitle


\section{Introduction}\label{sec:intro}

\begin{figure}
\includegraphics[width=0.5\textwidth]{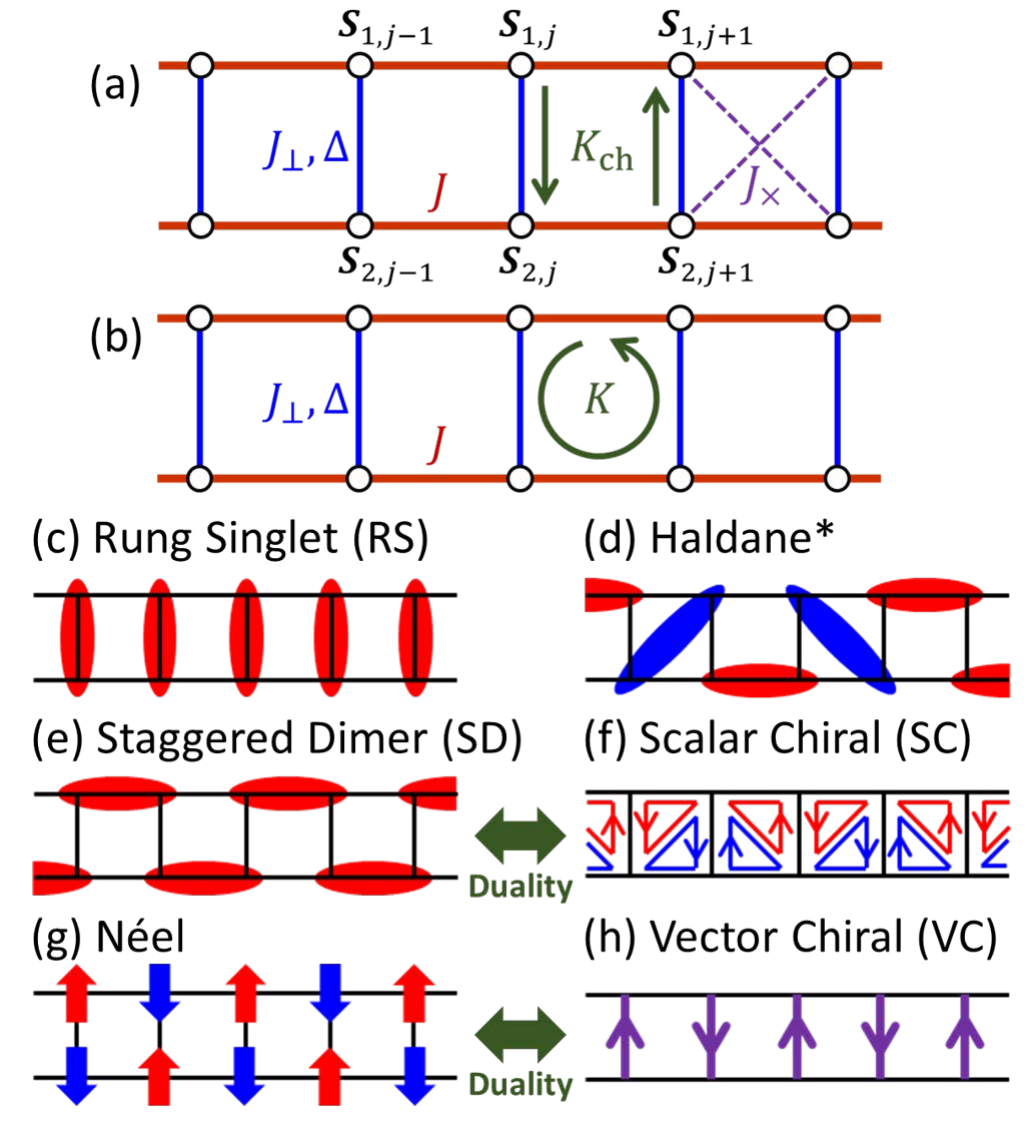}
\caption{\label{fig:laddermodel}
Spin-$\frac12$ XXZ ladders with (a) a CCI $K_{\rm ch}$ and (b) four-spin ring exchange $K$, 
which are described by the Hamiltonians \eqref{eq:H_XXZ_CCI} and \eqref{eq:H_XXZ_K}, respectively. 
The spin-$\frac12$ operator at the $j$th site on the $n$th leg is denoted by $\Sv_{n,j}$. 
Diagonal exchange interactions $J_\times$ are also considered in the extended model \eqref{eq:H_chex}. 
(c,d) Featureless states. 
Red and blue ovals indicate a singlet pair $(\ket{\ua\da}-\ket{\da\ua})/\sqrt{2}$ and a twisted singlet pair $(\ket{\ua\da}+\ket{\da\ua})/\sqrt{2}$, respectively. 
The Haldane* state is a superposition of various states covered by intra-leg singlet pairs and inter-leg twisted singlet pairs. 
(e,f) Ordered states around the isotropic case $\Delta=1$. 
In the SC state, each directed triangle indicates a positive scalar chirality of three spins along it. 
(g,h) Ordered states in the easy-axis case $\Delta>1$. 
In the N\'eel state, each arrow indicates an ordered magnetic moment along the $z$ axis. 
In the VC state, each directed bond indicates a positive vector chirality along it. 
The order parameters that characterize the states in (e,f,g,h) are shown in Eqs.\ \eqref{eq:O_SD_Neel} and \eqref{eq:O_SC_VC}. 
}
\end{figure}

Spin-$\frac12$ ladder systems have attracted considerable attention over three decades 
for their rich ground-state properties and their diverse material realizations \cite{Dagotto96,Rice96,Dagotto99,Lecheminant20,Sompet22}. 
The simplest model for such systems is a Heisenberg model on a two-leg ladder with exchange interactions $J$ and $J_\perp$ along the legs and rungs, respectively. 
When the leg interaction $J$ is antiferromagnetic ($J>0$), this model exhibits distinct gapped phases depending on the sign of $J_\perp$: 
the rung singlet (RS) phase for $J_\perp>0$ and the Haldane phase for $J_\perp<0$ \cite{Strong92,Strong94,Shelton96,Barnes93,White94}. 
The RS phase can be understood from the limit $J_\perp\to\infty$, 
where the ground state factorizes into singlet pairs on the rungs [Fig.\ \ref{fig:laddermodel}(c)]. 
In the Haldane phase, effective spin-$1$ degrees of freedom emerge on the rungs, and collectively form a spin-$1$ Haldane state \cite{Haldane83PLA, Haldane83PRL}. 
Both the RS and Haldane phases are {\it featureless} in the sense that their bulk ground state is unique and does not spontaneously break any symmetry. 
Nevertheless, these phases are distinct and cannot continuously be connected to each other in the presence of a certain symmetry, 
e.g., the discrete spin rotation symmetry $D_2=\mathbb{Z}_2\times\mathbb{Z}_2$ or time reversal \cite{GuZC09, Pollmann10, Pollmann12, ChenX11, Ogata20, Tasaki20book}; 
here, $D_2$ is generated by $\pi$ rotation of spins about two orthogonal axes. 
The Haldane phase can thus be viewed as a symmetry-protected topological (SPT) phase \cite{GuZC09, ChenX10, ChenX12, ChenX13, Zeng19book, Wen19Science, Senthil15} 
which is distinct from a rung-factorized product state. 

More examples of distinct featureless phases on a ladder can be found in the presence of exchange anisotropy. 
Liu {\it et al.}\  \cite{LiuZX12} have conducted a detailed classification of SPT phases in a spin-$\frac12$ two-leg ladder for the symmetry group $D_2\times\sigma$, 
where $\sigma$ is the symmetry with respect to the interchange of the two legs; 
this symmetry is {\it local} in the sense that each symmetry transformation can be decomposed into a product of unitary transformations acting on different rungs.  
They have found three new SPT phases termed $t_\mu~(\mu=x,y,z)$, which can be viewed as twisted variants of the Haldane phase. 
Specifically, the $t_\mu$ phase is related with the Haldane phase (termed $t_0$ in Ref.\ \cite{LiuZX12}) 
under the $\pi$ rotation of spins about the $\mu$-axis on the first leg, 
\begin{equation}\label{eq:U1mu}
U_1^\mu(\pi):=\exp \qty(i\pi \sum_j S_{1,j}^\mu). 
\end{equation}
These phases appear in highly anisotropic XXZ and XYZ ladder models \cite{LiuZX12, Manmana13, Fuji15, Fuji16}. 
Henceforth, we use the term ``the Haldane* phase'' \cite{Fuji15,Ogino21spt,Ogino22} to refer to the $t_z$ phase, which appears in the presence of XXZ anisotropy. 
The Haldane* state is a superposition of various states covered by intra-leg singlet pairs and inter-leg twisted singlet pairs, 
as shown in Fig.\ \ref{fig:laddermodel}(d). 

\begin{figure}
\includegraphics[width=0.47\textwidth]{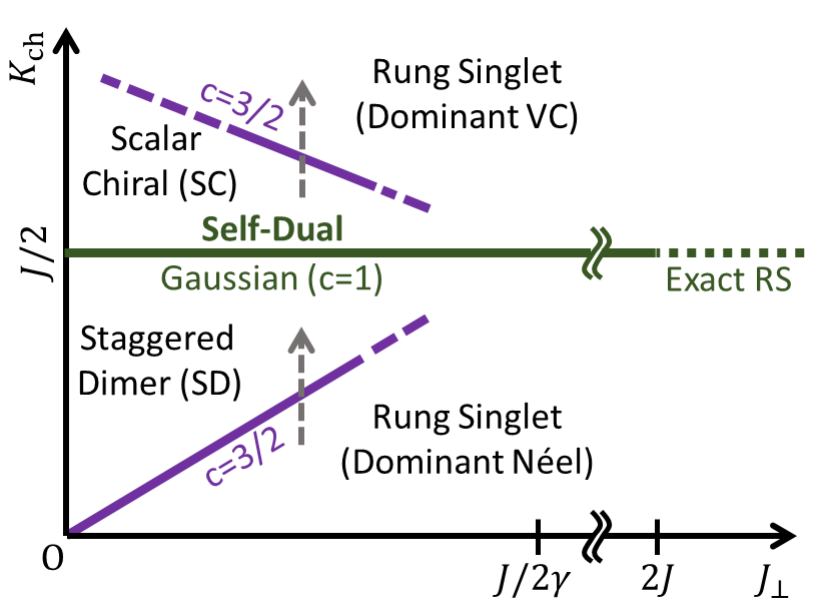}
\caption{\label{fig:phases_ch}
Schematic ground-state phase diagram of the XXZ-CCI ladder \eqref{eq:H_XXZ_CCI} in the isotropic case $\Delta=1$, 
predicted by the Abelian bosonization and the duality. 
This diagram is obtained by combining Fig.\ \ref{fig:phases_chex}(a,c) and setting $\Jdiag=0$. 
Along the two gray dashed arrows, we expect to find the phase structures in Fig.\ \ref{fig:phases_chex}(b,d) 
in the presence of exchange anisotropy ($\Delta\ne 1$). 
The self-dual line is located at $K_{\rm ch}=J/2$. 
Along this line and for $\Jperp>2J$, the ground state is exactly a product state of singlet pairs on the rungs; 
see Fig.\ \ref{fig:phases_selfdual} for more details. 
The SD-SC transition on the self-dual line belongs to the $c=1$ Gaussian universality class, 
according to a field-theoretical analysis around an SU$(4)$ quantum multicritical point \cite{Lecheminant05, Lecheminant06}. 
}
\end{figure}

\begin{figure*}
\includegraphics[width=0.8\textwidth]{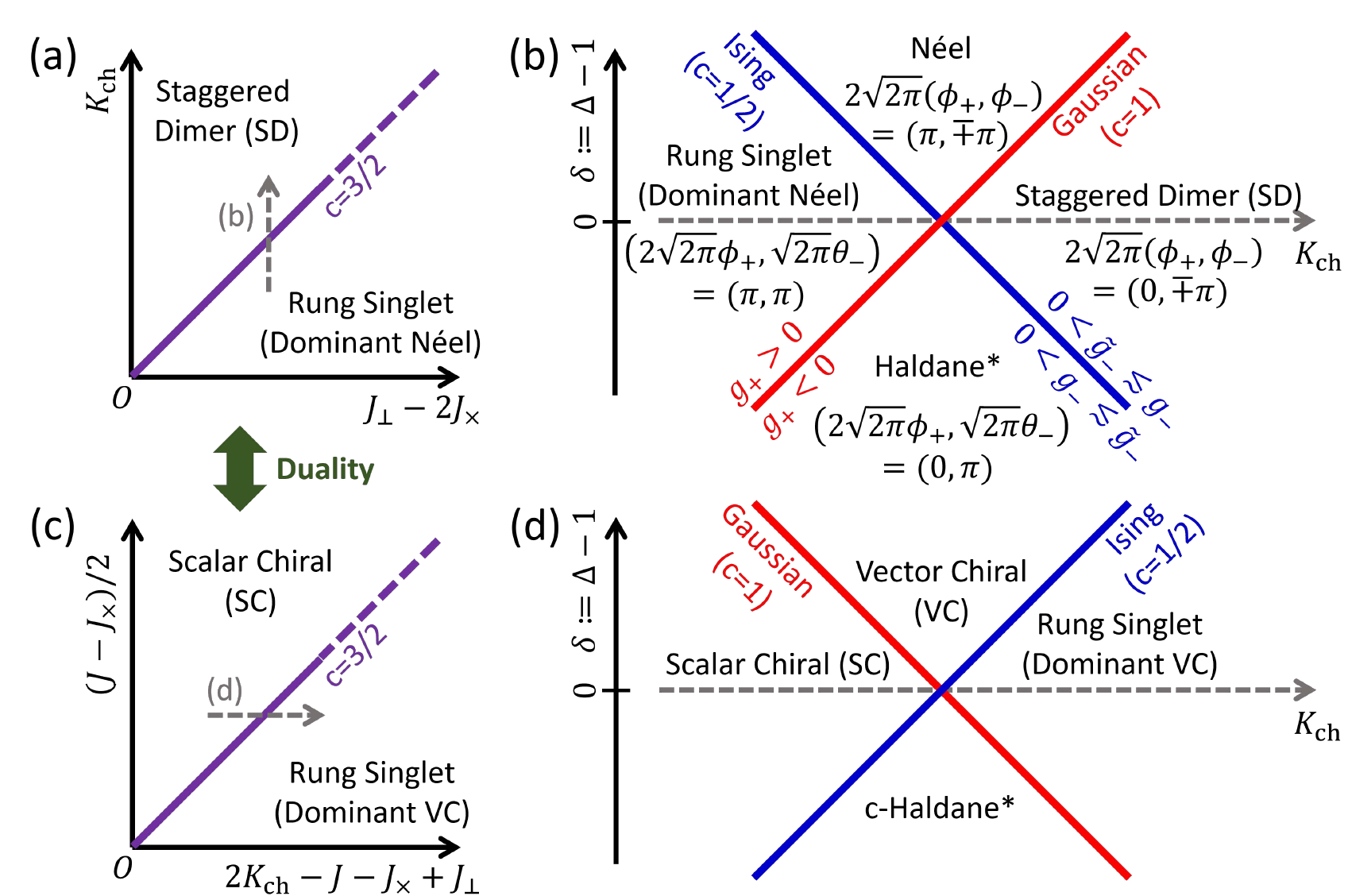}
\caption{\label{fig:phases_chex}
Schematic ground-state phase diagrams of the extended XXZ-CCI ladder \eqref{eq:H_chex}, 
predicted by the Abelian bosonization and the duality. 
(a) Isotropic model ($\Delta=1$) with weak inter-chain couplings \eqref{eq:H_chex_weak}. 
(c) Dual counterpart of (a), obtained through the transformation \eqref{eq:H_chex_duality}. 
(b,d) Anisotropic model $\Delta\ne 1$ along the dashed lines in (a,c). 
Each phase in (b) is characterized by the locking positions of bosonic fields; for more details, see Refs.\ \cite{Ogino21nvbs,Ogino21spt}. 
The transition lines in (b,d) are given by Eqs.\ \eqref{eq:Kch_Gauss_Ising} and \eqref{eq:Kch_Gauss_Ising_dual}, respectively; 
those in (a,c) are obtained by setting $\delta=0$ in these equations. 
}
\end{figure*}

Meanwhile, various types of four-spin interactions on a ladder have been studied as a route to exotic quantum phases. 
A natural extension of a Heisenberg ladder is to add four-spin ring exchange 
that appears in the fourth-order perturbation theory of the Hubbard model; 
its effects on excitation spectra have been discussed for La$_6$Ca$_8$Cu$_{24}$O$_{41}$ \cite{Brehmer99, Matsuda00}. 
A Heisenberg ladder with four-spin ring exchange $K$ [Fig.\ \ref{fig:laddermodel}(b) with $\Delta=1$] has been found to exhibit a remarkably rich phase diagram, 
which includes a staggered dimer (SD) phase [Fig.\ \ref{fig:laddermodel}(e)], a scalar chiral (SC) phase [Fig.\ \ref{fig:laddermodel}(f)], 
and a gapped featureless phase with dominant vector chirality correlation 
\cite{Laeuchli03, Hikihara03, Mueller02, Hijii02, Hijii03, Totsuka12, Furukawa06, Furukawa07, Metavitsiadis17, Hikihara08}. 
An exact spin-chirality duality provides a key to understanding this remarkable phase diagram \cite{Hikihara03, Momoi03, Hikihara08, Lecheminant05, Lecheminant06, Totsuka12}. 
Namely, the SC phase is dual to the SD phase, and the SD-SC transition line is located precisely on the self-dual point $K=J/2$. 
Furthermore, the dominant vector chirality phase is dual to the RS phase of the antiferromagnetic Heisenberg ladder which has dominant N\'eel-type spin correlation; 
in fact, these two phases are continuously connected to each other in the regime of strong $\Jperp>0$ \cite{Lecheminant06} 
(see Fig.\ \ref{fig:phases_ch} for an analogous situation). 
A Heisenberg ladder with an inter-leg dimer-dimer interaction (DDI) has also been studied 
\cite{Nersesyan97, LiYQ98, Pati98, Itoi00, Azaria00, Hijii09, Takayoshi10, Robinson19}
for its relevance to phonon-mediated interactions \cite{Nersesyan97} and spin-orbital systems \cite{LiYQ98,Pati98,Itoi00} as well as its similarity to a ring-exchange model. 
This model shows a SD phase as well as a gapless phase \cite{Pati98,Itoi00,Azaria00}. 
For both ring exchange ($K$) and an inter-leg DDI ($J_{\rm LL}$), field-theoretical analyses 
for weak interchain couplings ($|\Jperp|,|K|,|J_{\rm LL}|\ll J$) \cite{Mueller02, Nersesyan97, Takayoshi10, Robinson19} 
suggest that the RS-SD transition is continuous and described by the SU$(2)_2$ Wess-Zumino-Witten theory with the central charge $c=3/2$, 
although a possibility of a first-order transition is not excluded. 
Exact diagonalization results \cite{Hijii02, Hijii03, Hijii09} are consistent with the $c=3/2$ scenario for certain parameter ranges. 
In recent works, Ogino {\it et al.}\ \cite{Ogino21nvbs,Ogino21spt} have found that the presence of weak XXZ anisotropy has a significant impact on this transition. 
The RS-SD transition point with $c=3/2$ in the isotropic model has turned out to be a crossing point of Gaussian ($c=1$) and Ising ($c=1/2$) transition lines in the XXZ model, 
and the N\'eel phase [Fig.\ \ref{fig:laddermodel}(g)] and the Haldane* phase appear between these lines; see Fig.\ \ref{fig:phases_chex}(b). 
It is remarkable that the Haldane* phase, which has a highly anisotropic nature, appears in the vicinity of the isotropic case. 

In this paper, we aim to further explore combined effects of a four-spin interaction and XXZ anisotropy in a spin-$\frac12$ two-leg ladder. 
Specifically, we study an XXZ ladder with a chirality-chirality interaction (CCI) [Fig.\ \ref{fig:laddermodel}(a)], which is described by the Hamiltonian
\begin{equation}\label{eq:H_XXZ_CCI}
\begin{split}
 H
 &=J \sum_{n=1}^2 \sum_j \Sv_{n,j}\cdot\Sv_{n,j+1} 
 +\Jperp \sum_j \qty( \Sv_{1,j}\cdot\Sv_{2,j} )_\Delta \\
 &~~~~+K_{\rm ch} H_{\rm ch},
\end{split}
\end{equation}
where 
\begin{align}
 &\qty( \Sv_{n,i}\cdot\Sv_{m,j} )_\Delta := S_{n,i}^x S_{m,j}^x + S_{n,i}^y S_{m,j}^y + \Delta S_{n,i}^z S_{m,j}^z,\\
 &H_{\rm ch}=\sum_j 4 \qty( \Sv_{1,j}\times\Sv_{2,j} )\cdot\qty( \Sv_{1,j+1}\times\Sv_{2,j+1} ). \label{eq:H_CCI}
\end{align}
Henceforth, we refer to this model as the XXZ-CCI ladder. 
Throughout this paper, we focus on the case of $J,\Jperp,K_{\rm ch}>0$ unless stated otherwise. 
We set $J=1$ as the unit of energy when presenting numerical results. 
CCIs similar to Eq.\ \eqref{eq:H_CCI} have been studied as phonon-mediated interactions in multiferroics \cite{Onoda07, Furukawa08}. 

The CCI \eqref{eq:H_CCI} has close relationship with other four-spin interactions studied previously. 
Firstly, the CCI \eqref{eq:H_CCI} can be viewed as part of the ring exchange 
studied in Refs.\ \cite{Brehmer99, Matsuda00, Laeuchli03, Hikihara03, Mueller02, Hijii02, Hijii03, Furukawa06, Furukawa07, Hikihara08, Totsuka12, Metavitsiadis17}. 
Indeed, the ring exchange on a plaquette formed by the $j$th and $(j+1)$th rungs [$K$ in Fig.\ \ref{fig:laddermodel}(b)] is written as 
\begin{equation}\label{eq:P4_S}
\begin{split}
 &P_4(j,j+1)+P_4^{-1}(j,j+1)\\
 &=4\qty( \Sv_{1,j}\times\Sv_{2,j} )\cdot\qty( \Sv_{1,j+1}\times\Sv_{2,j+1} )+P_2(j)P_2(j+1)\\
 &~~~+\qty( \Sv_{1,j}+\Sv_{2,j} )\cdot\qty( \Sv_{1,j+1}+\Sv_{2,j+1} ), 
\end{split}
\end{equation}
where $P_2(j):=2\Sv_{1,j}\cdot\Sv_{2,j}+1/2$ is the two-spin exchange on the $j$th rung. 
As it turns out, the CCI induces similar phases as the ring exchange, as shown in Fig.\ \ref{fig:phases_ch}. 
However, the simple form of the CCI \eqref{eq:H_CCI} in terms of chirality can accentuate properties of the chirality-related phases, 
leading to more stable numerical results. 
It also simplifies analytical treatments as we will see later. 
Secondly, the CCI \eqref{eq:H_CCI} can also be written as 
\begin{equation}\label{eq:CCI_DDI}
 H_{\rm ch}=4H_{\rm LL}-4H_{\rm DD}
\end{equation}
with 
\begin{subequations}
\begin{align}
H_{\rm LL}=\sum_j \qty(\Sv_{1,j}\cdot\Sv_{1,j+1}) \qty(\Sv_{2,j}\cdot\Sv_{2,j+1}),\\
H_{\rm DD}=\sum_j \qty(\Sv_{1,j}\cdot\Sv_{2,j+1}) \qty(\Sv_{2,j}\cdot\Sv_{1,j+1}).
\end{align}
\end{subequations}
The first term $4H_{\rm LL}$ coincides with the inter-leg DDI studied 
in Refs.\ \cite{Nersesyan97, LiYQ98, Pati98, Itoi00, Azaria00, Hijii09, Takayoshi10, Robinson19, Ogino21nvbs,Ogino21spt}. 
According to field-theoretical analyses for weak interchain couplings \cite{Mueller02, Metavitsiadis17}, 
the second term $-4H_{\rm DD}$ only gives a small modification to the effect of the first term $4H_{\rm LL}$; see Sec.\ \ref{sec:bos} for more details. 
We therefore obtain a similar phase structure as that found by Ogino {\it et al.} \cite{Ogino21nvbs,Ogino21spt} in the presence of XXZ anisotropy; see Fig.\ \ref{fig:phases_chex}(b). 
However, a crucial difference occurs in the regime of strong four-spin interactions. 
There, the inter-leg DDI induces a gapless phase \cite{Pati98, LiYQ98, Itoi00, Azaria00, affleck1986exact, affleck1988critical} in the isotropic case, 
as opposed to gapped phases induced by a CCI in Fig.\ \ref{fig:phases_ch}. 

Our main results for the XXZ-CCI ladder \eqref{eq:H_XXZ_CCI} are schematically shown in Figs.\ \ref{fig:phases_ch} and \ref{fig:phases_chex}. 
Here, the ground-state phase diagram in the isotropic case $\Delta=1$ is shown in Fig.\ \ref{fig:phases_ch} 
while the phase structures around the $c=3/2$ critical points in the presence of XXZ anisotropy are shown in Fig.\ \ref{fig:phases_chex}(b,d). 
These phase diagrams are based on a field-theoretical analysis for weak interchain couplings as well as the spin-chirality duality. 
Our major contribution is on the phase structure around the SC-RS transition in Fig.\ \ref{fig:phases_chex}(d), 
which is a dual counterpart of the phase structure found by Ogino {\it et al.}\ \cite{Ogino21nvbs,Ogino21spt}. 
Here, the vector chiral (VC) phase is dual to the N\'eel phase, 
and exhibits a staggered pattern of vector chirality on the rungs, as shown in Fig.\ \ref{fig:laddermodel}(h). 
In contrast to the dominant vector chirality phase found in Refs.\ \cite{Laeuchli03, Hikihara03,Totsuka12}, 
the VC phase in this paper exhibits a genuine long-range order in terms of vector chirality, 
which is prohibited in the isotropic case but here allowed by XXZ anisotropy. 
We note that a different proposal to obtain a VC long-range order by the effects of a four-spin interaction and a magnetic field 
has been made in Refs.\ \cite{SatoM07VC, Totsuka12}; this has yet to be confirmed numerically. 
The c-Haldane* phase in Fig.\ \ref{fig:phases_chex}(d), where ``c'' stands for ``complex'' or ``chiral'',  is the dual counterpart of the Haldane* phase. 
To our knowledge, this is a new example of a SPT phase in the presence of the $D_2\times\sigma$ and time-reversal symmetries. 
In the XXZ-CCI ladder \eqref{eq:H_XXZ_CCI}, an exact self-dual surface exits at $K_{\rm ch}=J/2$. 
To reveal the phase structure and criticality around the self-dual surface, 
we apply the effective spinor hard-core bosons approach of Refs.\ \cite{Lecheminant06,Totsuka12} to the present model. 
We argue that the N\'eel-VC and Haldane*-c-Haldane* transitions occur in the easy-axis and easy-plane regimes, respectively, on the self-dual surface, 
and they both belong to the Gaussian universality class with $c=1$; see Figs.\ \ref{fig:phases_selfdual} and \ref{fig:phases_ch_Ising} shown later. 
We perform numerical analyses based on 
infinite density-matrix renormalization group (iDMRG) \cite{McCulloch07,McCulloch08, itensor,itensor-r0.3} 
and exact diagonalization \cite{QuSpin17, QuSpin19} 
to confirm the predicted phase structures and critical properties. 
We further demonstrate that the RS, Haldane*, and c-Haldane* phases are distinguished 
by topological indices associated with the $D_2\times\sigma$ and time-reversal symmetries. 

The rest of this paper is organized as follows. 
In Sec.\ \ref{sec:EFT}, we analyze the XXZ-CCI ladder \eqref{eq:H_XXZ_CCI} by applying the Abelian bosonization and the duality, 
and derive the schematic phase diagrams in Figs.\ \ref{fig:phases_ch} and \ref{fig:phases_chex}. 
In Sec.\ \ref{sec:bos_lat}, we present an effective description in terms of (spinor) hard-core bosons. 
In Secs.\ \ref{sec:numerics_isotropic} and \ref{sec:numerics_anisotropic}, we perform numerical analyses for the isotropic and anisotropic cases, respectively. 
In Sec.\ \ref{sec:summary}, we present a summary of this study and an outlook for future studies. 

\section{Field-theoretical analysis}\label{sec:EFT}

In this section, we analyze the ground-state phase diagram of the XXZ-CCI ladder \eqref{eq:H_XXZ_CCI} 
by means of the Abelian bosonization and the spin-chirality duality. 
For this purpose, we slightly extend the model by adding diagonal exchange couplings [Fig.\ \ref{fig:laddermodel}(a)] as 
\begin{equation}\label{eq:H_chex}
\begin{split}
 H_\text{ex}&=H
 +\Jdiag \sum_j \qty( \Sv_{1,j}\cdot\Sv_{2,j+1} + \Sv_{2,j}\cdot\Sv_{1,j+1} ).
\end{split}
\end{equation}
Although we set $\Jdiag=0$ at the end of the analysis, 
the role of the duality becomes clearer in this extended XXZ-CCI ladder \eqref{eq:H_chex}. 
As we explain later, the duality transformation can be described 
as a simple mapping of coupling constants in the extended model \eqref{eq:H_chex}. 

We start by analyzing the regime \footnote{
Here, we multiply $K_{\rm ch}$ by $2$ on the left-hand side for the following reason. 
Later, we will relax the strict weak-coupling condition, and assume that the bosonization results qualitatively continue 
even when the inequality \eqref{eq:H_chex_weak} is nearly saturated. 
There is a clear cutoff $K_{\rm ch}=J/2$ for this approach, which corresponds to the self-dual line in the absence of $\Jdiag$: see Fig.\ \ref{fig:phases_ch}. 
}
\begin{equation}\label{eq:H_chex_weak}
 \Jperp, |\Jdiag|, 2K_{\rm ch} \ll J,
\end{equation}
where the model \eqref{eq:H_chex} can be viewed as weakly coupled Heisenberg chains. 
In this regime, we can apply the Abelian bosonization \cite{Giamarchi04,Gogolin04} to formulate an effective field theory. 
Our formulation is an extension of those in Refs.\ \cite{Strong92,Strong94,Shelton96,Nersesyan97,Kim00,Mueller02,Vekua06,Takayoshi10,Robinson19,Ogino21nvbs,Ogino21spt,Ogino22}. 
We take notations similar to those in Refs.\ \cite{Takayoshi10,Ogino21nvbs,Ogino21spt,Ogino22} so that we can omit some details of the formalism. 
After analyzing the regime of Eq.\ \eqref{eq:H_chex_weak}, we apply the spin-chirality duality to discuss the regime of strong $K_{\rm ch}$. 

\subsection{Abelian bosonization for $\Jperp, |\Jdiag|, 2K_{\rm ch} \ll J$}\label{sec:bos}

In the limit $\Jpd/J , K_{\rm ch}/J\to 0$, each isolated chain labeled by $n=1,2$ is described 
by the Tomonaga-Luttinger liquid (TLL) theory (the Gaussian theory) 
in terms of a dual pair of bosonic fields $\phi_n(x)$ and $\theta_n(x)$. 
The spin and dimer operators on the $n$th chain are related to these bosonic fields as 
\begin{subequations}\label{eq:Spin_dim_bos}
\begin{align}
&S_{n, j}^z = \frac{a}{\sqrt{\pi}} \partial_x \phi_n + (-1)^j a_1 \cos(2\sqrt{\pi} \phi_n) + \cdots, \label{eq:Sz_bos}\\
&S_{n,j}^+ = e^{i\sqrt{\pi}\theta_n} \qty[b_0 (-1)^j + b_1 \cos(2\sqrt{\pi}\phi_n) + \cdots],\label{eq:Sp_bos}\\
&S_{n,j}^\alpha S_{n,j+1}^\beta=(-1)^j \delta_{\alpha\beta} d_\alpha \sin(2\sqrt{\pi}\phi_n) + \cdots,
\end{align}
\end{subequations}
where $S_{n,j}^\pm:=S_{n,j}^x\pm iS_{n,j}^y$, $a$ is the lattice constant, and $a_1$, $b_0$, $b_1$, and $d_\alpha~(\alpha=x,y,z)$ are non-universal coefficients 
\cite{LUKYANOV1997571,Hikihara98,Hikihara17,Takayoshi10}. 
In the presence of the SU$(2)$ symmetry, these coefficients satisfy $a_1 = b_0 =: \bar{a}$ and $d_x=d_y=d_z$. 

Treating the interchain $\Jpd$ and $K_{\rm ch}$ couplings perturbatively and expressing them using Eq.\ \eqref{eq:Spin_dim_bos}, 
we obtain the low-energy effective Hamiltonian of the extended XXZ-CCI ladder \eqref{eq:H_chex} as 
\begin{equation}\label{eq:H_chex_eff}
\begin{split}
H_{\rm ex}^{\rm eff}
&=\int \dd x \sum_{\nu=\pm} \frac{v_\nu}{2} \qty[ \frac1{K_\nu} \qty(\partial_x\phi_\nu) ^2+ K_\nu \qty(\partial_x\theta_\nu)^2] \\
&+ g_+ \cos(2\sqrt{2\pi}\phi_+) + g_- \cos(2\sqrt{2\pi}\phi_-) \\
&+ \tilde{g}_- \cos(\sqrt{2\pi}\theta_-)+\dots, 
\end{split}
\end{equation}
where
\begin{equation}\label{eq:phi_theta_pm}
\phi_{\pm} := \frac{1}{\sqrt{2}} (\phi_1 \pm \phi_2), ~~
\theta_{\pm} := \frac{1}{\sqrt{2}} (\theta_1 \pm \theta_2), 
\end{equation}
and ellipses indicate terms that have higher scaling dimensions. 
The bare coupling constants are given by  
\begin{subequations}\label{eq:coeff_cos}
\begin{align}
 g_\pm &= \frac{1}{2a} \qty[ a_1^2 \qty(\Jperp\Delta - 2\Jdiag) \mp 4K_{\rm ch}(9d^2-3d'^2) ], \label{eq:coeff_cos_gpm}\\
 \tilde{g}_-&=\frac{b_0^2}{a}  (\Jperp-2\Jdiag), \label{eq:coeff_gtm}
\end{align}
\end{subequations}
where $3d:=\sum_\alpha d_\alpha$ and $3d'^2:=\sum_{\alpha} d_\alpha^2$. 
These coupling constants are similar to those in a closely related model \cite{Ogino21nvbs,Ogino21spt}, 
but are slightly modified by the effects of the diagonal coupling $\Jdiag$ and the $-4H_{\rm DD}$ term in Eq.\ \eqref{eq:CCI_DDI}. 
The velocities $v_\pm$ and the TLL parameters $K_\pm$ in the symmetric and antisymmetric channels 
are in general modified from $v=\pi Ja/2$ and $K=1/2$ in the decoupled Heisenberg chains by the effects of the inter-chain couplings. 
The $g_\pm$ and $\tilde{g}_-$ terms have the scaling dimensions $2K_\pm$ and $(2K_-)^{-1}$, 
respectively, which are all equal to unity in the limit of decoupled Heisenberg chains where $K_\pm=1/2$. 
These terms are thus strongly relevant perturbations, and their competition dominantly determines the phase structure, as we discuss next. 

\subsection{Phase diagram from bosonization}\label{sec:bos_phases}

The effective Hamiltonian \eqref{eq:H_chex_eff}, in which the symmetric and antisymmetric channels are decoupled, 
can be analyzed in the same way as in Refs.\ \cite{Takayoshi10, Ogino21nvbs,Ogino21spt,Ogino22}. 
In the symmetric channel described by the sine-Gordon model, 
the strongly relevant $g_+$ term locks $\phi_+$ at distinct positions depending on the sign of $g_+$; 
a $c=1$ Gaussian transition occurs at $g_+=0$. 
The antisymmetric channel is described by the dual-field double sine-Gordon model. 
In this model, the strongly relevant $g_- $ and $\tilde{g}_-$ terms with close scaling dimensions (equal to unity in the limit of decoupled Heisenberg chains) crucially compete. 
A simple method to determine the phase structure is to examine which of $|g_-|$ and $|\tilde{g}_-|$ is larger. 
Specifically, $|g_-|\gtrsim |\tilde{g}_-|$ ($|g_-|\lesssim |\tilde{g}_-|$) leads to the locking of $\phi_-$ ($\theta_-$), 
and an $c=1/2$ Ising transition occurs at $|g_-|\approx |\tilde{g}_-|$ \cite{Shelton96,LECHEMINANT2002502}. 

Based on the above analysis and using the coupling constants in Eq.\ \eqref{eq:coeff_cos}, 
we obtain the schematic phase diagrams in Fig.\ \ref{fig:phases_chex}(a,b). 
In Fig.\ \ref{fig:phases_chex}(b), there are a Gaussian transition line (red) with 
\begin{subequations}\label{eq:Kch_Gauss_Ising}
\begin{equation}\label{eq:Kch_Gauss}
 K_{\rm ch} = \gamma \qty(\Jperp- 2\Jdiag+\Jperp \delta )
\end{equation}
as well as an Ising transition line (blue) with
\begin{equation}\label{eq:Kch_Ising}
 K_{\rm ch} = \gamma \qty(\Jperp- 2\Jdiag-\Jperp \delta ),
\end{equation}
\end{subequations}
where $\delta:=\Delta-1$. The proportionality constant $\gamma$ is estimated to be 
\begin{equation}\label{eq:gamma_est}
 \gamma= \frac{\bar{a}^2}{4(9d^2-3d'^2)}\approx 0.75,
\end{equation}
where we used the numerical values of $\bar{a}$, $d$, and $d'$ in the Heisenberg chain at a certain scale \cite{Takayoshi10}. 
The two transition lines cross at $J_4= \gamma (\Jperp-2\Jdiag)$ at the isotropic point $\Delta=1$, 
where the central charge is expected to be $c=3/2$ 
\cite{Nersesyan97, Hijii09, Hijii13, Takayoshi10, Robinson19, Mueller02, Tsuchiizu02}. 
Just with an infinitesimal anisotropy $\delta\ne0$, the N\'eel and Haldane* phases appear around this point. 

The SD and N\'eel phases are characterized by the order parameters
\begin{subequations}\label{eq:O_SD_Neel}
\begin{align}
\expval{\mathcal{O}_{\text{SD}}(j)} &= \frac14 
\langle \bm{S}_{1,j-1}\cdot\bm{S}_{1,j} - \bm{S}_{2,j-1}\cdot\bm{S}_{2,j} \notag\\
&~~~~~~-\bm{S}_{1,j}\cdot\bm{S}_{1,j+1} + \bm{S}_{2,j}\cdot\bm{S}_{2,j+1} \rangle, \label{eq:O_SD}\\
\expval{\mathcal{O}_{\text{N\'eel}}(j)} &= \frac14 \expval{S^z_{1,j} - S^z_{2,j} - S^z_{1,j+1} + S^z_{2,j+1}}. \label{eq:O_Neel}
\end{align}
\end{subequations}
See Fig.\ \ref{fig:laddermodel}(e,g). 
By using Eq.\ \eqref{eq:Spin_dim_bos}, these order parameters can be expressed in terms of the bosonic fields \eqref{eq:phi_theta_pm} as
\begin{subequations}\label{eq:O_SD_Neel_bos}
\begin{align}
 \expval{\mathcal{O}_{\text{SD}}(j)} 
 &=-(-1)^j (3d) \expval{\cos \qty(\sqrt{2\pi}\phi_+) \sin \qty(\sqrt{2\pi}\phi_-)}, \\
 \expval{\mathcal{O}_{\text{N\'eel}}(j)} 
 &= -(-1)^j a_1 \expval{\sin \qty(\sqrt{2\pi}\phi_+) \sin \qty(\sqrt{2\pi}\phi_-)}. 
\end{align}
\end{subequations}
We can easily see that these order parameters acquire finite values for the locking positions of the bosonic fields in Fig.\ \ref{fig:phases_chex}(b). 
For example, in the SD phase with $2\sqrt{2\pi}(\phi_+,\phi_-)=(0,\mp \pi)$, 
we have $\expval{\mathcal{O}_{\text{SD}}(j)}=\pm(-1)^j c_\text{SD}$, where $c_\text{SD}$ is a constant independent of $j$. 
For the characterizations of the RS and Haldane* phases in bosonization, see, e.g., Refs.\ \cite{Fuji15,Ogino21nvbs,Ogino21spt}. 

\subsection{Spin-chirality duality}\label{sec:duality}

The spin-chirality duality \cite{Hikihara03, Momoi03, Lecheminant05, Lecheminant06, Totsuka12} 
can be described by the unitary transformation $\Ucal(\pi/2)$, whose explicit form is shown later in Eq.\ \eqref{eq:U_theta}. 
Under this transformation, the staggered spin component and the vector chirality on each rung are interchanged as 
\begin{subequations}\label{eq:U_spin_chirality}
\begin{align}
&\Ucal \qty(\frac{\pi}2) \qty(S_{1,j}^\alpha-S_{2,j}^\alpha) \Ucal^\dagger \qty(\frac{\pi}2) = -2 \qty(\Sv_{1,j}\times\Sv_{2,j})^\alpha,\\
&\Ucal \qty(\frac{\pi}2) \qty(\Sv_{1,j}\times\Sv_{2,j})^\alpha \Ucal^\dagger \qty(\frac{\pi}2) = \frac12 \qty(S_{1,j}^\alpha-S_{2,j}^\alpha),
\end{align}
\end{subequations}
while the uniform spin component $\Sv_{1,j}+\Sv_{2,j}$ is invariant. 
The SD and N\'eel order parameters in Eq.\ \eqref{eq:O_SD_Neel} are then transformed into the SC and VC order parameters 
\begin{subequations}\label{eq:O_SC_VC}
\begin{align}
\expval{\mathcal{O}_{\text{SC}}(j)} &= \frac14 
\langle \qty(\Sv_{1,j-1}+\Sv_{2,j-1})\cdot\qty(\Sv_{1,j}\times\Sv_{2,j}) \notag\\
&~~~~+\qty(\Sv_{1,j-1}\times\Sv_{2,j-1})\cdot\qty(\Sv_{1,j}+\Sv_{2,j}) \notag\\
&~~~~-\qty(\Sv_{1,j}+\Sv_{2,j})\cdot\qty(\Sv_{1,j+1}\times\Sv_{2,j+1}) \notag\\
&~~~~-\qty(\Sv_{1,j}\times\Sv_{2,j})\cdot\qty(\Sv_{1,j+1}+\Sv_{2,j+1})\rangle, \label{eq:O_SC}\\
\expval{\mathcal{O}_{\text{VC}}(j)} &= \frac12 \expval{\qty(\Sv_{1,j}\times\Sv_{2,j}-\Sv_{1,j+1}\times\Sv_{2,j+1})^z}, \label{eq:O_VC}
\end{align}
\end{subequations}
respectively, through the relations 
\begin{align}\label{eq:SD_SC_Neel_VC}
  \Ucal \qty(\frac{\pi}2) \mathcal{O}_{\text{SD}}(j) \Ucal^\dagger \qty(\frac{\pi}2) &= -\mathcal{O}_{\text{SC}}(j), \\
  \Ucal \qty(\frac{\pi}2) \mathcal{O}_{\text{N\'eel}}(j) \Ucal^\dagger \qty(\frac{\pi}2) &= -\mathcal{O}_{\text{VC}}(j).
\end{align}
We can thus obtain SC and VC phases as dual counterparts of the SD and N\'eel phases, respectively; see Fig.\ \ref{fig:laddermodel}(e,f,g,h). 

To apply the duality, it is convenient to rewrite the Hamiltonian \eqref{eq:H_chex} of the extended XXZ-CCI ladder as \cite{Lecheminant06,Totsuka12}
\begin{equation}\label{eq:H_chex_H12ch}
\begin{split}
 H_{\rm ex}&=\frac{J+\Jdiag}{2} H_1+\frac{J-\Jdiag}{2}H_2+K_{\rm ch} H_{\rm ch}\\
 &~~~+\Jperp \sum_j \qty(\Sv_{1,j}\cdot\Sv_{2,j})_\Delta,
\end{split}
\end{equation}
where 
\begin{subequations}
\begin{align}
 H_1&=\sum_j \qty( \Sv_{1,j}+\Sv_{2,j} )\cdot\qty( \Sv_{1,j+1}+\Sv_{2,j+1} ),\\
 H_2&=\sum_j \qty( \Sv_{1,j}-\Sv_{2,j} )\cdot\qty( \Sv_{1,j+1}-\Sv_{2,j+1} ). 
\end{align}
\end{subequations}
Under the duality transformation \eqref{eq:U_spin_chirality}, 
$H_2$ and $H_{\rm ch}$ are interchanged while the other terms in Eq.\ \eqref{eq:H_chex_H12ch} remain invariant. 
Therefore, the duality transformation of the Hamiltonian can be described as the simple exchange of the coupling constants,
\begin{equation}\label{eq:H_chex_duality}
 \frac{J-\Jdiag}{2} \leftrightarrow K_{\rm ch}. 
\end{equation}
Here, the other coupling constants $(J+\Jdiag)/2$ and $\Jperp$ in Eq.\ \eqref{eq:H_chex_H12ch} remain unchanged. 

\subsection{Full phase structure}\label{sec:phases}

Under the duality transformation, the phase diagram of the isotropic model in Fig.\ \ref{fig:phases_chex}(a) is mapped to the one in Fig.\ \ref{fig:phases_chex}(c). 
Here, the SD phase is mapped to the SC phase as explained above. 
The RS phase with dominant N\'eel correlation is mapped to the RS phase with dominant VC correlation. 
The condition of weak inter-chain couplings in Eq.\ \eqref{eq:H_chex_weak} is now replaced by the following condition: 
\begin{equation}\label{eq:condition_dual}
 \Jperp, \bigg|\frac{J+\Jdiag}{2} -K_{\rm ch}\bigg|,J-\Jdiag \ll \frac{J+\Jdiag}{2}+K_{\rm ch}.
\end{equation} 

When the anisotropy $\Delta\ne 1$ is introduced along the dashed arrow in Fig.\ \ref{fig:phases_chex}(c), 
we expect to find the phase structure in Fig.\ \ref{fig:phases_chex}(d). 
Here, the VC and c-Haldane* phases are the dual counterparts of the N\'eel and Haldane* phases, respectively. 
In this diagram, there are a Gaussian transition line (red) with
\begin{subequations}\label{eq:Kch_Gauss_Ising_dual}
\begin{equation}\label{eq:Kch_Gauss_dual}
 K_{\rm ch} = \frac{J+\Jdiag-\Jperp}{2} + \frac{J-\Jdiag}{4\gamma}- \frac{\Jperp\delta}{2} 
\end{equation}
as well as an Ising transition line (blue) with 
\begin{equation}\label{eq:Kch_Ising_dual}
 K_{\rm ch} = \frac{J+\Jdiag-\Jperp}{2} + \frac{J-\Jdiag}{4\gamma}+ \frac{\Jperp\delta}{2} ,
\end{equation}
\end{subequations}
which are obtained from Eq.\ \eqref{eq:Kch_Gauss_Ising} via the transformation \eqref{eq:H_chex_duality}. 
To our knowledge, the c-Haldane* phase is a new example of a SPT phase in the presence of the $D_2\times\sigma$ and time-reversal symmetries. 
We will discuss the Haldane* and c-Haldane* phases and their distinction in more detail in Secs.\ \ref{sec:EasyPlane} and \ref{sec:topo} and Appendix \ref{app:topo_MPS}. 

By combining Fig.\ \ref{fig:phases_chex}(a,c) and setting $\Jdiag=0$, we obtain the phase diagram in Fig.\ \ref{fig:phases_ch}. 
Here, we assumed that the phase structures in Fig.\ \ref{fig:phases_chex}(a,c) qualitatively continue 
up to the regime where the conditions \eqref{eq:H_chex_weak} and \eqref{eq:condition_dual} are nearly saturated. 
This assumption is justified {\it a posteriori} through an overall agreement 
between the predicted phase diagram in Fig.\ \ref{fig:phases_ch} and the numerically obtained one in Fig.\ \ref{fig:phases_ch_ed}. 
We also note that predictions of bosonization made in weak-coupling regimes often hold qualitatively beyond their validity ranges; 
see, e.g., Refs.\ \cite{Robinson19,Hikihara10}. 
Along the two gray dashed arrows in Fig.\ \ref{fig:phases_ch}, we expect to find the phase structures in Fig.\ \ref{fig:phases_chex}(b,d) 
in the presence of exchange anisotropy ($\Delta\ne 1$). 
The SD-SC transition is located exactly on the self-dual line with $K_{\rm ch}=J/2$; 
this belongs to the $c=1$ Gaussian universality class, 
according to a field-theoretical analysis around an SU$(4)$ quantum multicritical point \cite{Lecheminant05, Lecheminant06}. 


Lastly, we discuss the method of the polarization amplitude
\cite{Resta98, RestaSorella99, AligiaOrtiz99, AligiaHallberg00, NakamuraTodo02, NakamuraVoit02, Nakamura03, FuruyaSato21, Tasaki18}, 
which is useful for numerically determining some of the phase boundaries. 
In a $2$-leg ladder of finite length $L$, the polarization amplitude is defined as 
\begin{equation}\label{eq:z_rung}
 z_{\rm rung} = \Bigg\langle \exp\qty( i\frac{2\pi}{L} \sum_{n=1}^2 \sum_{j=1}^L j S_{n,j}^z) \Bigg\rangle , 
\end{equation}
which has the form of the expectation value of the Lieb-Schultz-Mattis twist operator \cite{Lieb61,Oshikawa97}. 
In the Abelian bosonization formulation explained in Secs.\ \ref{sec:bos} and \ref{sec:bos_phases}, 
the polarization amplitude has the expression \cite{NakamuraVoit02, Nakamura03, FuruyaSato21} 
\begin{equation}\label{eq:z_rung_phi}
 z_{\rm rung}\propto -\Big\langle \cos\qty(2\sqrt{2\pi} \phi_+) \Big\rangle . 
\end{equation}
It therefore changes its sign when the locking position of $2\sqrt{2\pi} \phi_+$ changes across the Gaussian transition as in Fig.\ \ref{fig:phases_chex}(b). 
As Eq.\ \eqref{eq:z_rung} is invariant under the duality transformation, 
it can be used to detect the Gaussian transition in Fig.\ \ref{fig:phases_chex}(d) as well. 
In fact, the polarization amplitude \eqref{eq:z_rung} has been used to detect the RS-SD and SC-RS transitions in the Heisenberg ladder with ring exchange \cite{Laeuchli03}. 
We will use the polarization amplitude \eqref{eq:z_rung} to analyze the regime around the isotropic case in Secs.\ \ref{sec:numerics_isotropic} and \ref{sec:numerics_anisotropic}.

\subsection{Implications for an XXZ ladder with four-spin ring exchange}\label{sec:XXZ_K}

By means of the Abelian bosonization and the duality, 
we have analyzed the phase structure of the XXZ-CCI ladder \eqref{eq:H_XXZ_CCI} and its extension \eqref{eq:H_chex}. 
Here, we briefly discuss implications of this analysis 
for an XXZ ladder with four-spin ring exchange [Fig.\ \ref{fig:laddermodel}(b)], which is described by the Hamiltonian 
\begin{equation}\label{eq:H_XXZ_K}
\begin{split}
 H_\text{XXZ-$K$}=&J \sum_{n=1}^2 \sum_j \Sv_{n,j}\cdot\Sv_{n,j+1} +\Jperp \sum_j \qty( \Sv_{1,j}\cdot\Sv_{2,j} )_\Delta\\
 &~~+K\sum_j \qty[ P_4(j,j+1)+P_4^{-1}(j,j+1) ]. 
\end{split}
\end{equation}
Here, the ring exchange term is expressed in terms of spin operators as Eq.\ \eqref{eq:P4_S}. 
Compared to the model studied in Refs.\ \cite{Laeuchli03, Hikihara03, Mueller02, Hijii02, Hijii03, Totsuka12}, 
XXZ anisotropy is introduced in the $\Jperp$ couplings on the rungs. 
Henceforth we refer to this model \eqref{eq:H_XXZ_K} as the XXZ-$K$ ladder. 
Below we focus on the case of $J,\Jperp,K>0$. 

To highlight the difference from the XXZ-CCI ladder, it is useful to rewrite Eq.\ \eqref{eq:H_XXZ_K} as 
\begin{equation}\label{eq:H_XXZ_K_H11chRR}
\begin{split}
 H_\text{XXZ-$K$} &= \qty(\frac{J}{2}+K) H_1+\frac{J}{2} H_2 + K H_{\rm ch} +4KH_{\rm RR}\\
 &~~~+\Jperp \sum_j \qty(\Sv_{1,j}\cdot\Sv_{2,j})_\Delta + 2K \sum_j \Sv_{1,j}\cdot\Sv_{2,j}\\
 &~~~+{\rm const.},
\end{split}
\end{equation}
where 
\begin{equation}
 H_{\rm RR}=\sum_j \qty( \Sv_{1,j}\cdot\Sv_{2,j} ) \qty(\Sv_{1,j+1}\cdot\Sv_{2,j+1}).
\end{equation}
Compared to a similar expression \eqref{eq:H_chex_H12ch} for the extended XXZ-CCI ladder, 
a crucial difference is the presence of the $H_{\rm RR}$ term. 
In the Abelian bosonization analysis for weak inter-chain couplings $\Jperp, 2K\ll J$, 
the $H_{\rm RR}$ term only has the effect of removing the ``$3d'^2$'' term in Eq.\ \eqref{eq:coeff_cos_gpm}; 
namely, a precise cancellation of relevant terms occurs between the $H_{\rm RR}$ and $H_{\rm DD}$ terms \cite{Mueller02, Metavitsiadis17} 
while relevant terms from the two-spin interactions and the $H_{\rm LL}$ term remain.  
We therefore expect a phase structure similar to Fig.\ \ref{fig:phases_chex}(a,b). 
For large $K$, however, the analysis is nontrivial. 
Under the duality transformation \eqref{eq:U_spin_chirality}, $H_2$ and $H_{\rm ch}$ are interchanged 
while the other terms in Eq.\ \eqref{eq:H_XXZ_K_H11chRR} remain invariant; 
the self-dual surface is thus located at $K=J/2$. 
In contrast to $H_{\rm ch}$, 
$H_{\rm RR}$ remains to have the same complicated form under the duality transformation. 
When its coefficient $4K$ becomes large, it is nontrivial to predict the effects of $H_{\rm RR}$ on the phase structure 
by some analytical calculations. 
Yet, the DMRG studies for $\Jperp=J$ and $\Delta=1$ \cite{Laeuchli03, Hikihara03} have identified the RS-SD-SC-RS structure of the phase diagram, 
similar to a constant-$\Jperp$ cut of Fig.\ \ref{fig:phases_ch}. 
When the XXZ anisotropy $\Delta\ne 1$ is introduced, we thus expect to find the phase structre in Fig.\ \ref{fig:phases_chex}(d) around the SC-RS transition. 
We leave the determination of the phase diagram of the XXZ-$K$ ladder \eqref{eq:H_XXZ_K} for a future work. 
In our preliminary results, we have found that compared to the XXZ-CCI ladder, the presence of $H_{\rm RR}$ 
tends to obscure the properties of chirality-related phases in numerical simulations and require a more careful analysis. 
Thus, our study on the XXZ-CCI ladder \eqref{eq:H_XXZ_CCI} in this paper will provide a useful guide to analyze the XXZ-$K$ ladder \eqref{eq:H_XXZ_K}. 

\section{Effective spin-$1$ hard-core bosons approach}\label{sec:bos_lat}

In this section, we apply the effective spin-1 hard-core bosons approach of Refs.\ \cite{Lecheminant06,Totsuka12} to the XXZ-CCI ladder \eqref{eq:H_XXZ_CCI}. 
This approach helps us determine the phase structures and the critical properties 
near the self-dual surface $K_{\rm ch}=J/2$ in the easy-axis and easy-plane regimes. 

\subsection{Spin-$1$ hard-core bosons}\label{sec:spinor_bosons}

To formulate the spin-$1$ bosons picture, it is convenient to introduce the singlet-triplet basis on each rung as 
\begin{subequations}\label{eq:stxyz}
\begin{align}
 \ket{s} &= \frac{1}{\sqrt{2}} \qty( \ket{\ua\da}- \ket{\da\ua}),\\
 \ket{t_x}&= -\frac{1}{\sqrt{2}} \qty( \ket{\ua\ua} - \ket{\da\da} ),\\
 \ket{t_y}&= \frac{i}{\sqrt{2}} \qty( \ket{\ua\ua} + \ket{\da\da} ),\\
 \ket{t_z}&= \frac{1}{\sqrt{2}} \qty( \ket{\ua\da} + \ket{\da\ua} ).
\end{align}
\end{subequations}
The creation operator for the spin-$1$ hard-core boson on the $j$th rung is then defined as 
\begin{equation}\label{eq:b_ts}
 b_{j,\alpha}^\dagger = |t_\alpha^{[j]}\rangle \langle s^{[j]}|~~(\alpha=x,y,z), 
\end{equation}
where the superscript $[j]$ indicates the concerned rung. 
This operator can also be expressed as a complex combination of the staggered spin component and the vector chirality, i.e., 
\begin{equation}\label{eq:b_S}
 b_{j,\alpha}^\dagger = \frac12 (\Sv_{1,j}-\Sv_{2,j})^\alpha+i(\Sv_{1,j}\times\Sv_{2,j})^\alpha. 
\end{equation}
The number of bosons on the $j$th rung is given by
\begin{equation}\label{eq:nB_bb}
 n^{\rm B}_j = \sum_{\alpha} b_{j,\alpha}^\dagger b_{j,\alpha} = \sum_{\alpha} |t_\alpha^{[j]}\rangle \langle t_\alpha^{[j]}|.
\end{equation}
As this is nothing but the projection operator onto the triplet sector on the rung, 
it has eigenvalues of zero and unity for the singlet and triplet states, respectively, signifying the hard-core nature of bosons. 

To discuss the duality, we introduce the unitary transformation
\begin{equation}\label{eq:U_theta}
 \Ucal (\theta) = \exp(i\theta \sum_j n^{\rm B}_j)~~(\theta\in\mathbb{R}),
\end{equation}
which plays a role of a gauge transformation for bosons as
\begin{equation}\label{eq:UbU}
 \Ucal (\theta) b_{j,\alpha}^\dagger \Ucal^\dagger (\theta) = e^{i\theta} b_{j,\alpha}^\dagger.
\end{equation}
The spin-chirality duality is given by $\Ucal(\pi/2)$. 
Indeed, by setting $\theta=\pi/2$, Eq.\ \eqref{eq:UbU} combined with Eq.\ \eqref{eq:b_S} gives Eq.\ \eqref{eq:U_spin_chirality}.

\subsection{Spinor bosons Hamiltonian}\label{sec:H_bos}

In the picture of spinor hard-core bosons, the Hamiltonian \eqref{eq:H_XXZ_CCI} of the XXZ-CCI ladder can be expressed as
\begin{equation}\label{eq:H_XXZ_CCI_bos}
\begin{split}
 H
 &= t\sum_{j,\alpha} \left(b_{j,\alpha}^\dagger b_{j+1,\alpha} + b_{j+1,\alpha}^\dagger b_{j,\alpha} \right)\\
 &~+u\sum_{j,\alpha} \left(b_{j,\alpha}^\dagger b_{j+1,\alpha}^\dagger + b_{j+1,\alpha} b_{j,\alpha} \right)
 +J_{\rm bl} \sum_j \Tv_j\cdot\Tv_{j+1}\\
 &~+\Jperp \sum_j \qty[ n_j^{\rm B} -\frac34 + \frac{\delta}{2}\qty(b_{j,x}^\dagger b_{j,x} + b_{j,y}^\dagger b_{j,y} -\frac12) ] \\
 &~+\text{const.},
\end{split}
\end{equation}
where $\delta:=\Delta-1$ and 
\begin{equation}\label{eq:tuJ}
 t=\frac{J}{2}+K_{\rm ch},~~
 u=\frac{J}{2}-K_{\rm ch},~~
 J_{\rm bl}=\frac{J}{2}. 
\end{equation}
Here, $\Tv_j$ is the spin of the hard-core boson on the $j$th rung, which can be expressed as 
\begin{equation}
 T^\alpha_j = -i\epsilon_{\alpha\beta\gamma} b_{j\beta}^\dagger b_{j\gamma}. 
\end{equation}
Under the duality transformation $\Ucal(\pi/2)$, the sign of $u$ is flipped ($u\to -u$) while the other coupling constants in Eq.\ \eqref{eq:H_XXZ_CCI_bos} remain unchanged. 
The Hamiltonian \eqref{eq:H_XXZ_CCI_bos} has the global U$(1)$ gauge symmetry under the transformation \eqref{eq:U_theta} with arbitrary $\theta$ in the self-dual case $u=0$, 
and explicitly breaks it otherwise. 
The self-dual model with $u=0$ is analogous to spin-$1$ bosonic atoms loaded into an optical lattice \cite{Demler02, Imambekov03, Kawaguchi12, StamperKurn13}. 
In this analogy, the $\Jperp$ term in Eq.\ \eqref{eq:H_XXZ_CCI_bos} plays a role of spin-dependent chemical potentials. 

The global U$(1)$ gauge symmetry of the self-dual model with $u=0$ allows us to apply the Lieb-Schultz-Mattis-type theorem \cite{Tasaki20book,Lieb61,Oshikawa97}. 
We introduce the number of bosons per rung as 
\begin{equation}\label{eq:nB}
 n^{\rm B}:=\frac1L \sum_{j=1}^L n_j^{\rm B},
\end{equation}
where $L$ is the ladder length, i.e., the total number of rungs. 
According to the theorem, a unique bulk ground state below an excitation gap is allowed only for $n^{\rm B}=0$ and $1$. 
The case of $n^{\rm B}=0$ (i.e., the vacuum of bosons) corresponds to the exact RS ground state \cite{Mueller02,Kolezhuk98}, 
and is realized for sufficiently large antiferromagnetic $\Jperp>0$. 
The case of $n^{\rm B}=1$ corresponds to an effective spin-$1$ chain, and is realized for sufficiently large ferromagnetic $\Jperp<0$; 
there, gapped featureless phases such as the Haldane phase and a trivial (so-called, large-$D$) phase appear \cite{Schulz86, denNijs89,ChenHida03}. 
For $0<n^{\rm B}<1$, the system must show gapless excitations or degenerate ground states in the thermodynamic limit. 

\begin{figure}
\includegraphics[width=0.4\textwidth]{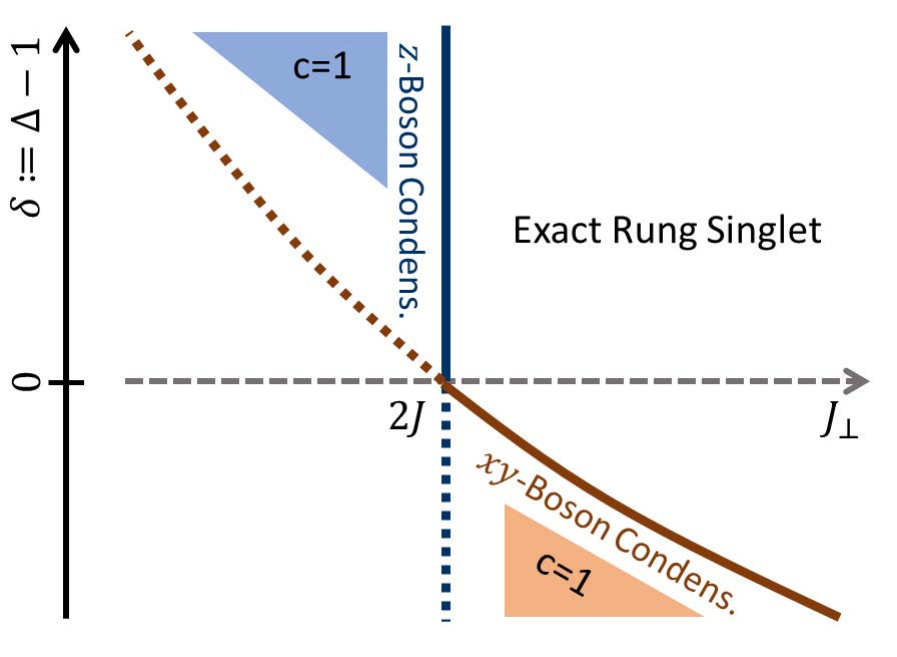}
\caption{\label{fig:phases_selfdual}
Ground-state phase diagram of the XXZ-CCI ladder \eqref{eq:H_XXZ_CCI} on the self-dual surface $K_{\rm ch}=J/2$, 
obtained from the spinor hard-core bosons Hamiltonian \eqref{eq:H_XXZ_CCI_bos} with $u=0$. 
For $\Jperp>2J {\rm max}(1,(1+\delta/2)^{-1})$, where all of Eq.\ \eqref{eq:epsilon_xyz} are gapped, 
the ground state is the boson vacuum, i.e., the exact RS state \cite{Mueller02,Kolezhuk98}. 
This ground state is destabilized when the minimum of $\epsilon_{x,y}(k)$ or $\epsilon_z(k)$ in Eq.\ \eqref{eq:epsilon_xyz} becomes zero, 
i.e., the $xy$- or $z$-bosons condense. 
The shaded $c=1$ region for easy-axis anisotropy ($\delta>0$) corresponds to the N\'eel-VC transition in Fig.\ \ref{fig:phases_ch_Ising}. 
The shaded $c=1$ region for easy-plane anisotropy ($\delta<0$) corresponds to the Haldane*-c-Haldane* transition. 
The region of the exact RS corresponds to a crossover line in the RS phase across which dominant correlation changes from N\'eel- to VC-type, 
as shown in Figs.\ \ref{fig:phases_ch} and \ref{fig:phases_ch_Ising}. 
}
\end{figure}

The range of the exact RS ground state in the self-dual model with $u=0$ can be determined 
by examining the single-boson excitation spectrum \cite{Mueller02,Kolezhuk98}. 
This spectrum can be determined easily and exactly from the Hamiltonian \eqref{eq:H_XXZ_CCI_bos} 
as the boson-boson interaction plays no role in the single-boson problem. 
The dispersion relations $\epsilon_\alpha(k)$ for the $\alpha$-boson ($\alpha=x,y,z$) in Eq.\ \eqref{eq:b_ts} are obtained as 
\begin{subequations}\label{eq:epsilon_xyz}
\begin{align}
 \epsilon_{x,y}(k) &=2J\cos k + \Jperp \qty(1+\frac{\delta}{2}),\\
 \epsilon_z(k) &=2J\cos k + \Jperp,
\end{align}
\end{subequations}
where $k$ is the wave number. 
These bosons are gapped for sufficiently large $\Jperp>0$, where the exact RS ground state is realized. 
This ground state is destabilized when the minimum $\epsilon_{x,y}(\pi)$ or $\epsilon_z(\pi)$ of the dispersion relations becomes zero. 
This analysis leads to the range of the exact RS state in Fig.\ \ref{fig:phases_selfdual}. 

\begin{figure}
\includegraphics[width=0.4\textwidth]{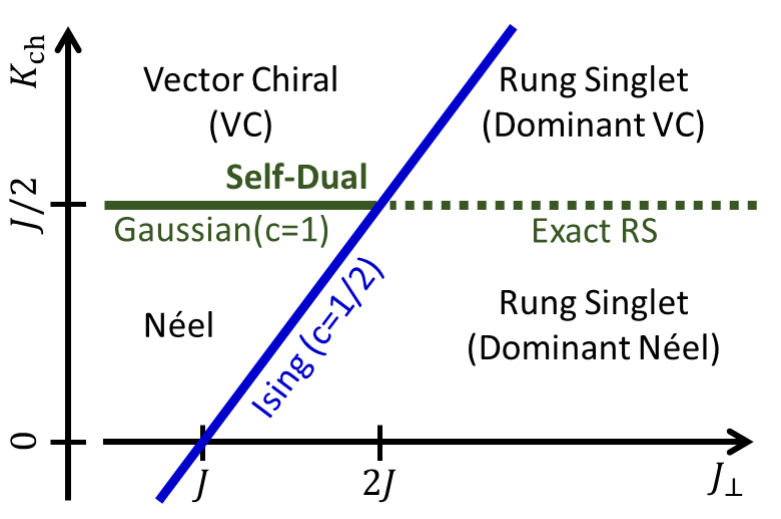}
\caption{\label{fig:phases_ch_Ising}
Ground-state phase diagram of the XXZ-CCI ladder \eqref{eq:H_XXZ_CCI} in the easy-axis regime, 
obtained from the effective XY Hamiltonian \eqref{eq:H_XY}. 
Here, we assume that $\Jperp \delta$ with $\delta:=\Delta-1$ is much larger than the other coupling constants. 
}
\end{figure}

\subsection{Easy-axis regime} \label{sec:EasyAsis}

\newcommand{\Scal}{{\cal S}}

When $\Jperp\delta\gg J,K_{\rm ch}$, 
the $xy$-bosons can safely be neglected in the bosonic Hamiltonian \eqref{eq:H_XXZ_CCI_bos}.
This can be seen in the dispersion relations \eqref{eq:epsilon_xyz}, where the $xy$-bosons have an energy difference of $\Jperp\delta/2$ from the $z$-boson. 
Focusing on the $z$-bosons and introducing the pseudospin-$\frac12$ operator 
\begin{subequations}\label{eq:Scal_bz}
\begin{align}
 \Scal^x_j &= \frac12 \qty(b_{j,z}^\dagger + b_{j,z}),\\
 \Scal^y_j &= \frac1{2i} \qty(b_{j,z}^\dagger - b_{j,z}),\\
 \Scal^z_j &= b_{j,z}^\dagger b_{j,z} - \frac12, 
\end{align}
\end{subequations}
we can rewrite Eq.\ \eqref{eq:H_XXZ_CCI_bos} as the Hamiltonian of the XY chain in a magnetic field 
\begin{equation}\label{eq:H_XY}
 H = \sum_j \qty( 2J\Scal_j^x \Scal_{j+1}^x + 4K_{\rm ch} \Scal_j^y \Scal_{j+1}^y ) + \Jperp \sum_j \Scal_j^z + \text{const.}.
\end{equation}
The ground-state phase diagram of this Hamiltonian is well-known \cite{Dutta15book, BarouchMcCoy71}. 
When $\Jperp>J+2K_{\rm ch}$, the system is in a ``paramagnetic phase'', 
in which pseudospins are mainly polarized in the $-z$ direction and no spontaneous symmetry breaking occurs. 
This corresponds to the RS phase of the ladder model. 
When $\Jperp<J+2K_{\rm ch}$, the system exhibits an antiferromagnetic order along the $x$ axis ($J>2K_{\rm ch}$) or the $y$ axis ($J<2K_{\rm ch}$). 
These correspond to the N\'eel and VC phases of the ladder model, 
as seen from the relations $\Scal_j^x=(S_{1,j}^z-S_{2,j}^z)/2$ and $\Scal_j^y=(\Sv_{1,j}\times\Sv_{2,j})^z$ obtained from Eqs.\ \eqref{eq:b_S} and \eqref{eq:Scal_bz}. 
This analysis leads to the phase diagram in Fig.\ \ref{fig:phases_ch_Ising}. 
Here, the RS-N\'eel and RS-VC transitions belong to the Ising universality class, 
which is consistent with the bosonization analysis in Fig.\ \ref{fig:phases_chex} that started from the different limits. 

\subsection{Easy-plane regime} \label{sec:EasyPlane}

\newcommand{\Lv}{\bm{L}}
\newcommand{\Lt}{\tilde{L}}
\newcommand{\Kch}{K_{\rm ch}}
\newcommand{\Brm}{{\rm B}}
\newcommand{\Frm}{{\rm F}}
\newcommand{\Jbl}{J_{\rm bl}}

When $-\Jperp\delta\gg J,K_{\rm ch}$, 
the $z$-boson can safely be neglected in the bosonic Hamiltonian \eqref{eq:H_XXZ_CCI_bos}. 
We introduce new bosonic operators 
\begin{equation}
 b_{j,\pm}^\dagger=\frac{1}{\sqrt{2}} \qty( \mp b_{j,x}^\dagger -ib_{j,y}^\dagger ) .
\end{equation}
As these can also be expressed as $b_{j,+}^\dagger=|\! \ua\ua^{[j]}\rangle \langle s^{[j]}|$ 
and $b_{j,-}^\dagger=|\! \da\da^{[j]}\rangle \langle s^{[j]}|$, 
these correspond to the creation of the states $\ket{\ua\ua}$ and $\ket{\da\da}$, respectively, on the $j$th rung. 
After removing the $z$-bosons, the Hamiltonian \eqref{eq:H_XXZ_CCI_bos} can be expressed in terms of the new bosonic operators as 
\begin{equation}\label{eq:H_bpm}
\begin{split}
 H
 &= t\sum_{\nu=\pm}\sum_j \left(b_{j,\nu}^\dagger b_{j+1,\nu} + b_{j+1,\nu}^\dagger b_{j,\nu} \right)\\
 &~~-u\sum_{\nu=\pm}\sum_j \left(b_{j,\nu}^\dagger b_{j+1,-\nu}^\dagger + b_{j+1,-\nu} b_{j,\nu} \right)\\
 &~~+J_{\rm bl} \sum_j T_j^z T_{j+1}^z+D\sum_{\nu=\pm} \sum_j b_{j,\nu}^\dagger b_{j,\nu} + \text{const.},
\end{split}
\end{equation}
where $D=\Jperp (1+\delta/2)$ and $T_j^z=\sum_{\nu=\pm} \nu b_{j,\nu}^\dagger b_{j,\nu}$. 
While $t$, $u$, and $J_{\rm bl}$ are mutually dependent as in Eq.\ \eqref{eq:tuJ}, we treat them as independent parameters 
in discussing the properties of the effective Hamiltonian \eqref{eq:H_bpm}. 
When $u=J_{\rm bl}=0$, Eq.\ \eqref{eq:H_bpm} can be viewed as a model of two-component hard-core bosons on a lattice 
with the hopping amplitude $t$ and the chemical potential $-D$. 
This model is equivalent to the infinite-$U$ limit of the fermionic Hubbard chain under a statistical transformation \cite{Takayoshi10SPI}. 
For the convenience of our analysis, we relax infinite $U$ to a large but finite intercomponent repulsion $0<U<\infty$ in the following. 
When $D<2t$, the finite-$U$ model shows the two-component TLL phase 
described by dual pairs of bosonic fields $(\phi_\rho,\theta_\rho)$ and $(\phi_\sigma,\theta_\sigma)$ in the charge and spin sectors, respectively \cite{Giamarchi04,Gogolin04}. 
We may then treat the $u$ and $\Jbl$ terms in Eq.\ \eqref{eq:H_bpm} perturbatively, 
which play roles of a pairing term and state-dependent density-density interactions, respectively. 
In this way, we obtain an effective field-theoretical Hamiltonian 
\begin{equation}\label{eq:H_charge_spin}
\begin{split}
 H^{\rm eff} 
 &= \int \dd x \sum_{\eta=\rho,\sigma} \frac{v_\eta}{2} 
  \qty[ \frac1{K_\eta} \qty(\partial_x\phi_\eta) ^2+ K_\eta \qty(\partial_x\theta_\eta)^2]\\
 &~~+g_\sigma\cos(2\sqrt{2\pi} \phi_\sigma) +\tilde{g}_\rho \cos( \sqrt{2\pi} \theta_\rho )+\dots,
\end{split}
\end{equation}
where $v_\rho$ and $K_\rho$ ($v_\sigma$ and $K_\sigma$) are the velocity and the TLL parameter in the charge (spin) sector, 
$\tilde{g}_\rho\propto u$, 
and ellipses indicate terms that have higher scaling dimensions. 
In the fermionic Hubbard chain with finite $U$ and $u=J_{\rm bl}=0$, we have $K_\sigma=1$ and $\tilde{g}_\rho=0$, 
and the $g_\sigma$ term with the scaling dimension $2K_\sigma$ is marginally irrelevant. 
In the infinite-$U$ limit, we have $K_\rho\to 1/2$ and $v_\sigma\to 0$ \cite{Giamarchi04}. 
Finite $\Jbl>0$ leads to the opening of a gap in the spin sector via the $g_\sigma$ term. 
We therefore have the $c=1$ gapless state of the charge sector on the self-dual surface with $u=0$, 
as shown in Fig.\ \ref{fig:phases_selfdual}. 
For $u\ne 0$, the $\tilde{g}_\rho$ term with the scaling dimension $(2K_\rho)^{-1}$ immediately opens a gap. 
In this way, we have distinct gapped phases depending on the sign of $u=\frac{J}{2}-\Kch$, 
and a $c=1$ Gaussian transition occurs between them at $u=0$. 

What are the gapped phases for $\frac{J}{2}>\Kch$ and $\frac{J}{2}<\Kch$? 
To answer this question, we rewrite Eq.\ \eqref{eq:H_bpm} as the Hamiltonian of a spin-$1$ chain 
by regarding the states 
\begin{equation}\label{eq:basis_Hals}
 \{ \ket{\ua\ua}, \ket{s}, -\ket{\da\da} \}
\end{equation}
on each rung $j$ as the basis for a spin-$1$ operator $\Lv_j$. 
Here, the states \eqref{eq:basis_Hals} are related to the usual triplet states $\{ \ket{\ua\ua}, \ket{t_z}, \ket{\da\da} \}$ 
on a rung via the transformation $(-i)^L U_1^z(\pi)$; see Eq.\ \eqref{eq:U1mu}. 
The spin-$1$ operators $L_j^+:=L_j^x+iL_j^y$ and $L_j^z$ are expressed in terms of the bosonic operators $b_{j,\pm}$ as 
\begin{subequations}
\begin{align}
 &L_j^+ = \sqrt{2} \qty(b_{j,+}^\dagger - b_{j,-}),\\
 &L_j^z = b_{j,+}^\dagger b_{j,+} - b_{j,-}^\dagger b_{j,-}. 
\end{align}
\end{subequations}
We also introduce 
\begin{equation}\label{eq:Lvt}
 \tilde{\Lv}_j = e^{i\frac{\pi}{2} n_j^{\rm B}}  \Lv_j e^{-i\frac{\pi}{2} n_j^{\rm B}} ,
\end{equation}
where $e^{i\frac{\pi}{2}n_j^{\rm B}}$ is the single-rung part of the duality transformation $\Ucal(\pi/2)$ defined by Eq.\ \eqref{eq:U_theta} with $\theta=\pi/2$. 
Equation \eqref{eq:Lvt} can be viewed as the spin-$1$ operator in the basis 
\begin{equation}\label{eq:basis_cHals}
 \{ -\ket{\ua\ua}, i\ket{s}, \ket{\da\da} \}. 
\end{equation}
which is related to Eq.\ \eqref{eq:basis_Hals} via $i^L\Ucal (\pi/2)$. 
We note that $\Lt_j^z=L_j^z$. 
Using $\Lv_j$ and $\tilde{\Lv}_j$, we obtain a spin-$1$ chain Hamiltonian as 
\begin{equation}\label{eq:H_L_Lt}
\begin{split}
 H
 &=\frac{J}{2} \sum_j \qty( L_j^x L_{j+1}^x + L_j^y L_{j+1}^y )\\
 &~~~+ \Kch \sum_j \qty( \Lt_j^x \Lt_{j+1}^x + \Lt_j^y \Lt_{j+1}^y ) \\
 &~~~+ \Jbl \sum_j L_j^z L_{j+1}^z
 +D\sum_j \qty(L_j^z)^2+\text{const.}. 
\end{split}
\end{equation} 
We again treat $J/2$, $\Kch$, and $\Jbl$ as independent parameters in discussing the properties of the Hamiltonian \eqref{eq:H_L_Lt}. 
When $\Kch=0$, Eq.\ \eqref{eq:H_L_Lt} is the spin-$1$ XXZ chain with single-ion anisotropy in terms of the $\Lv_j$ spins; 
see Refs.\ \cite{Schulz86, denNijs89, ChenHida03} for the phase diagram of this spin-$1$ chain. 
The Haldane state of the $\Lv_j$ spins corresponds to the Haldane* state (or the $t_z$ state) discussed in Refs.\ \cite{LiuZX12, Fuji15,Ogino21spt,Ogino22}. 
When $\frac{J}{2}=0$, Eq.\ \eqref{eq:H_L_Lt} is the spin-$1$ XXZ chain with single-ion anisotropy in terms of the $\tilde{\Lv}_j$ spins. 
The Haldane state of the $\tilde{\Lv}_j$ spins corresponds to the c-Haldane* state, 
which is the dual counterpart of the Haldane state. 
When both $\frac{J}{2}$ and $\Kch$ are finite, a competition between the Haldane* and c-Haldane* phases is expected. 
We thus conclude that gapped phases for $\frac{J}{2}>\Kch$ and $\frac{J}{2}<\Kch$ indicated by the field-theoretical analysis 
are the Haldane* and c-Haldane* phases, respectively, and a $c=1$ Gaussian transition occurs between them. 

\section{Numerical results: isotropic case}\label{sec:numerics_isotropic}

\begin{figure}
\includegraphics[width=0.42\textwidth]{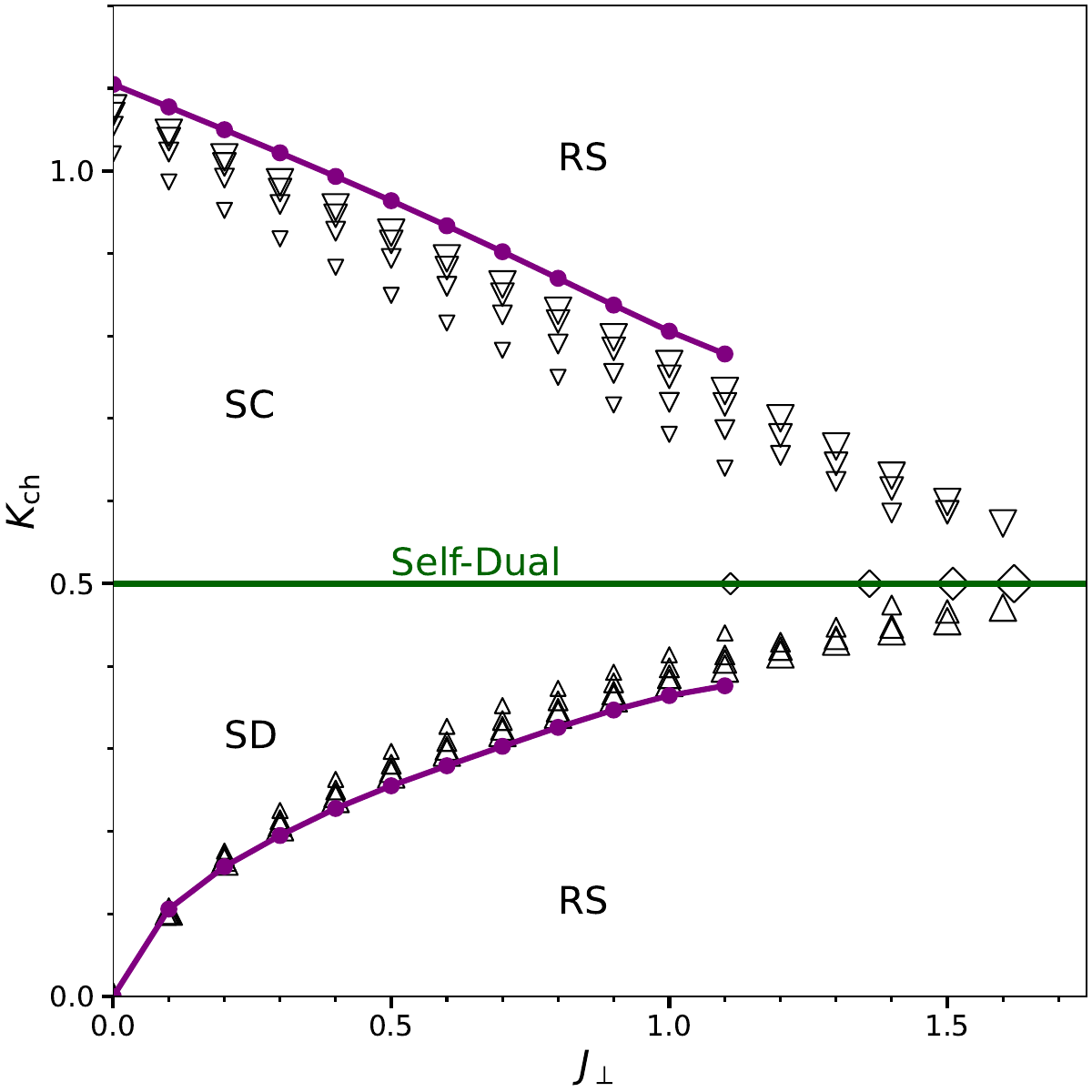} 
\caption{\label{fig:phases_ch_ed}
Ground-state phase diagram of the XXZ-CCI ladder \eqref{eq:H_XXZ_CCI} in the isotropic case $\Delta=1$, determined by exact diagonalization. 
Here, we set $J=1$ as the unit of energy. 
The self-dual line is located at $K_{\rm ch}=1/2$ (green), which gives the exact SD-SC transition line for $\Jperp<2$. 
Up- and down-pointing triangles show finite-size estimates of the RS-SD and SC-RS transition points, respectively; 
these are based on the sign changes of the polarization ampbiltude $z_{\rm rung}$, as shown in Fig.\ \ref{fig:J05_fit}.  
These triangles merge at rhombi on the self-dual line. 
Four different symbol sizes (from small to large) correspond to the ladder lengths $L=8$, $10$, $12$, and $14$; 
the number of spins in the ladder is $2L$ in total. 
The transition points extrapolated to the infinite-size limit are shown by purple filled circles interpolated by solid lines; 
this extrapolation is not done for $\Jperp\ge 1.2$, where the $L$-dependence of the estimated transition points is not smooth. 
See Fig.\ \ref{fig:phases_ch} for a schematic phase diagram predicted by the Abelian bosonization and the duality. 
}
\end{figure}

\begin{figure}
\includegraphics[width=0.45\textwidth]{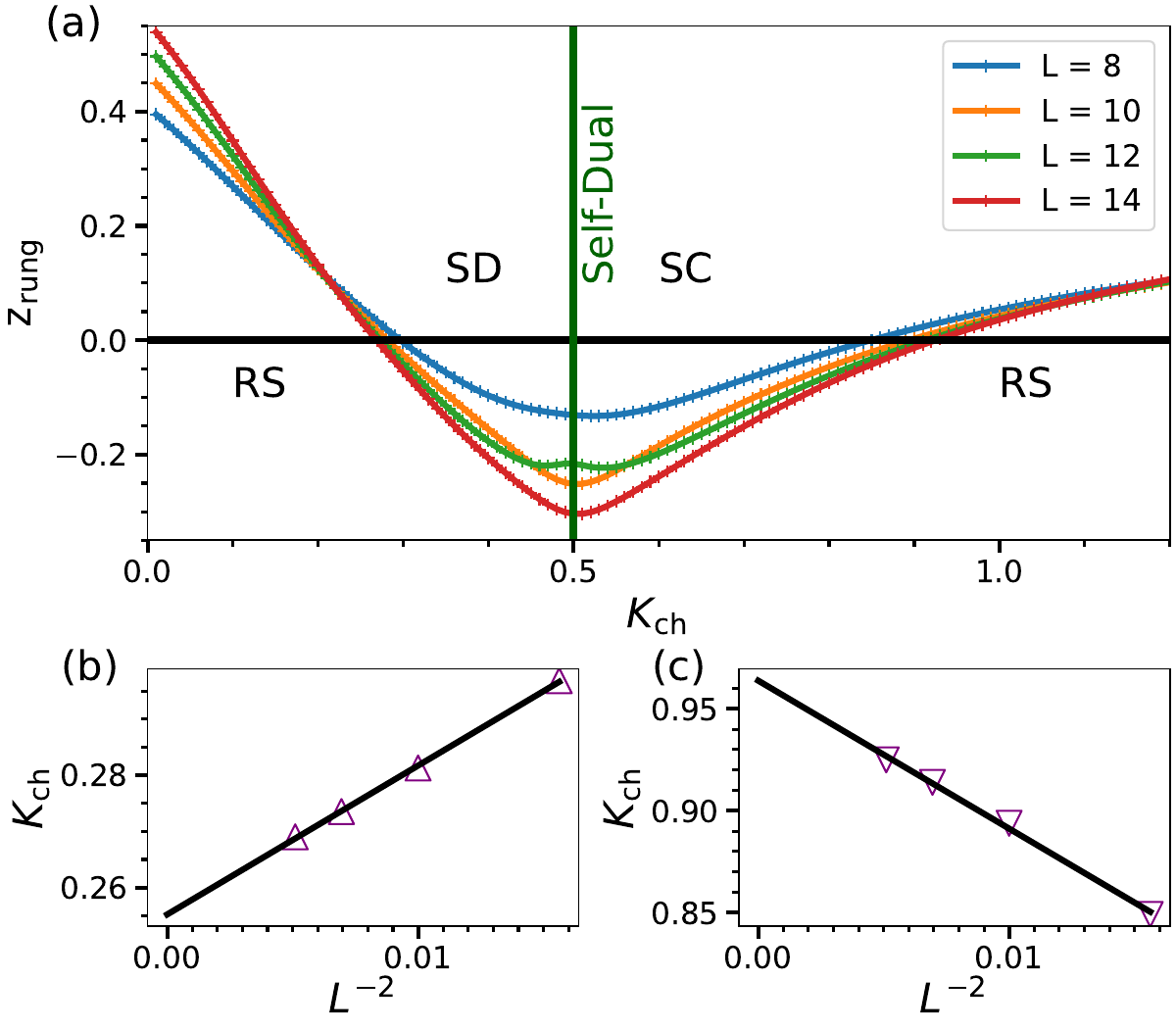}
\caption{\label{fig:J05_fit}
(a) Polarization amplitude $z_{\rm rung}$ in Eq.\ \eqref{eq:z_rung} as a function of $K_{\rm ch}$, 
obtained by exact diagonalization for the case of $\Jperp=0.5$ in Fig.\ \ref{fig:phases_ch_ed}. 
Results for different ladder lengths $L$ are shown by different colors. 
The sign changes of $z_{\rm rung}$ give finite-size estimates of the RS-SD and SC-RS transitions. 
(b,c) Extraplation of the RS-SD [(b)] and SC-RS [(c)] transition points, based on a linear function of $L^{-2}$ \cite{NakamuraTodo02}. 
The transition point in the infinite-size limit is given by the intercept. 
}
\end{figure}

In this section, we present numerical results on the XXZ-CCI ladder \eqref{eq:H_XXZ_CCI} in the isotropic case $\Delta=1$. 
Here, we set $J=1$ as the unit of energy. 
The obtained phase diagram is shown in Fig.\ \ref{fig:phases_ch_ed}, 
which is in overall agreement with the schematic phase diagram in Fig.\ \ref{fig:phases_ch} predicted by the Abelian bosonization and the duality. 

The phase diagram in Fig.\ \ref{fig:phases_ch_ed} is based on the exact diagonalization calculation of the polarization amplitude $z_{\rm rung}$ in Eq.\ \eqref{eq:z_rung}. 
Here, exact diagonalization was performed with the Python package QuSpin \cite{QuSpin17, QuSpin19}. 
Figure \ref{fig:J05_fit} shows an example of this analysis for $\Jperp=0.5$. 
Finite-size estimates of the RS-SD and SC-RS transition points are obtained through the sign changes of $z_{\rm rung}$, as shown in Fig.\ \ref{fig:J05_fit}(a). 
The estimated transition points are then extrapolated to the infinite-size limit by using a linear function of $L^{-2}$, as shown in Fig.\ \ref{fig:J05_fit}(b). 
This extrapolation procedure works well for $\Jperp\lesssim 1.1$, where the transition points are not too close to the self-dual line. 
For larger $\Jperp$, however, the $L$-dependence of the estimated transition points is not smooth, 
and thus we only present finite-size results in Fig.\ \ref{fig:phases_ch_ed}. 
We note that the polarization amplitude $z_{\rm rung}$ is insensitve to the SD-SC transition; 
however, the spin-chirailty duality as well as the hard-core bosons picture in Sec.\ \ref{sec:H_bos} (see Fig.\ \ref{fig:phases_selfdual}) 
indicate that it should be located on the self-dual line $K_{\rm ch}=1/2$ with $\Jperp<2$.  

\begin{figure}
\includegraphics[width=0.4\textwidth]{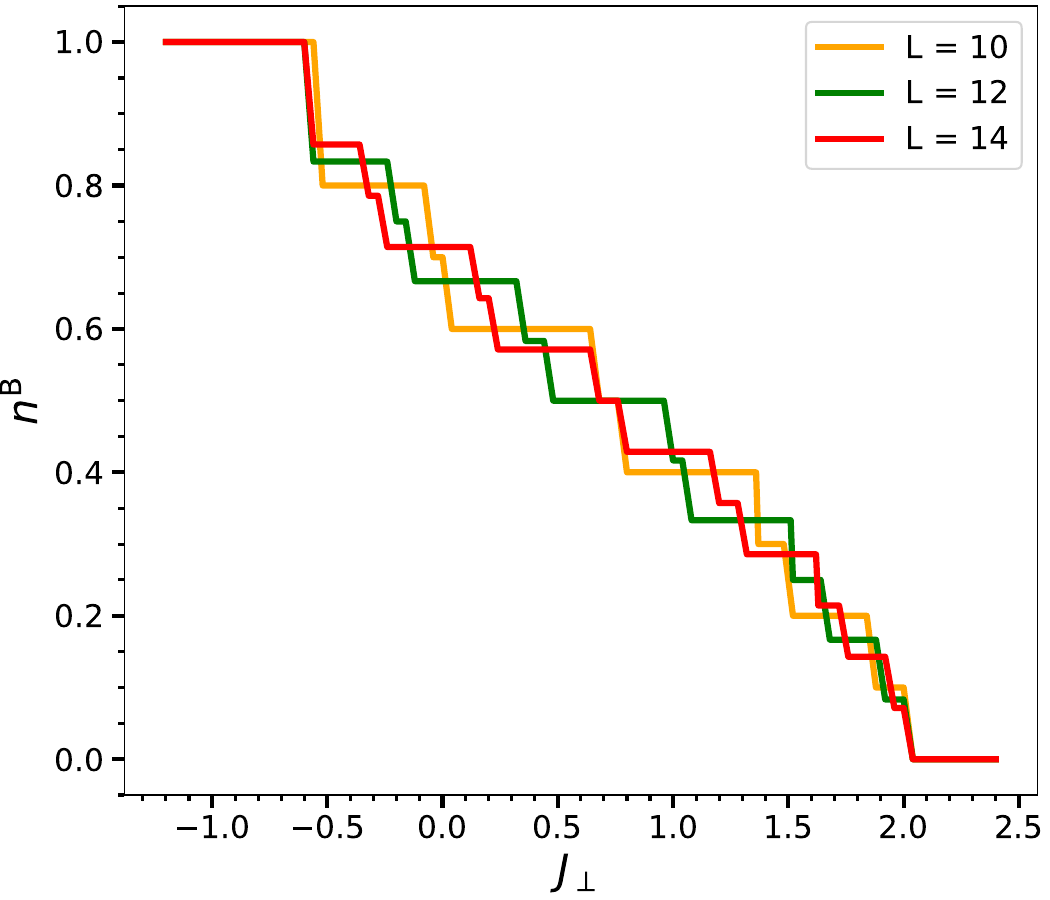} 
\caption{\label{fig:boson}
The ground-state value of the boson density $n^B$ in Eq.\ \eqref{eq:nB}, 
calculated by exact diagonalization along the self-dual line $K_{\rm ch}=1/2$ in Fig.\ \ref{fig:phases_ch_ed}. 
Results for different ladder lengths $L$ are shown by different colors. 
}
\end{figure}

In the hard-core bosons picture in Sec.\ \ref{sec:H_bos}, 
the total number of bosons is a good quantum number on the self-dual line $K_{\rm ch}=1/2$. 
Furthermore, in the isotropic case $\Delta=1$, the $\Jperp$ term plays a role of 
a chemical potential term for bosons that commutes with the Hamiltonian; see Eq.\ \eqref{eq:H_XXZ_CCI_bos}.
Therefore, in finite-size ladders, the ground-state value of the boson density $n^B$ in Eq.\ \eqref{eq:nB} changes discontinuously and stepwise as a function of $\Jperp$, 
in a way similar to magnetization curves in spin systems with the spin-rotational U$(1)$ symmetry. 
Figure \ref{fig:boson} shows this density $n^{\rm B}$ calculated by exact diagonalization. 
For $\Jperp>2$, where the chemical potential $-\Jperp$ is sufficiently low, bosons are fully depleted ($n^{\rm B}=0$); 
we therefore have the boson vacuum, i.e., the exact RS state \cite{Mueller02,Kolezhuk98}, in agreement with Fig.\ \ref{fig:phases_selfdual}. 
For $\Jperp\lesssim -0.6$,  hard-core bosons are fully filled ($n^{\rm B}=1$), 
and the system exhibits the Haldane ground state of the spin-$1$ Heisenberg chain that corresponds to the $J_{\rm bl}$ term in Eq.\ \eqref{eq:H_XXZ_CCI_bos}. 
For $-0.6\lesssim \Jperp<2$, the boson density $n^{\rm B}$ increases with lowering $\Jperp$. 
As discussed on the basis of the Lieb-Schultz-Mattis-type theorem in Sec.\ \ref{sec:H_bos}, 
the system must show gapless excitations or degenerate ground states in the thermodynamic limit when $0<n^{\rm B}<1$. 
In Fig.\ \ref{fig:boson}, we find no robust plateau-like feature in the boson density $n^{\rm B}(\ne 0,1)$ that is associated with an energy gap. 
We thus expect that the scenario of gapless excitations holds on the self-dual line with $-0.6\lesssim \Jperp<2$. 
This implies that the SD-SC transition that occurs on this line is continuous. 
When the system deviates from the self-dual line, the stepwise features of the boson density in Fig.\ \ref{fig:boson} are gradually smoothed. 
Near the self-dual line, the remaining stepwise features of $n^{\rm B}$ can lead to nontrivial finite-size effects in other quantities as well; 
this explains the difficulty in extrapolating the transition points for $\Jperp\ge 1.2$ in Fig.\ \ref{fig:phases_ch_ed}. 

\section{Numerical results: anisotropic case}\label{sec:numerics_anisotropic}

\begin{figure*}
\includegraphics[width=0.7\textwidth]{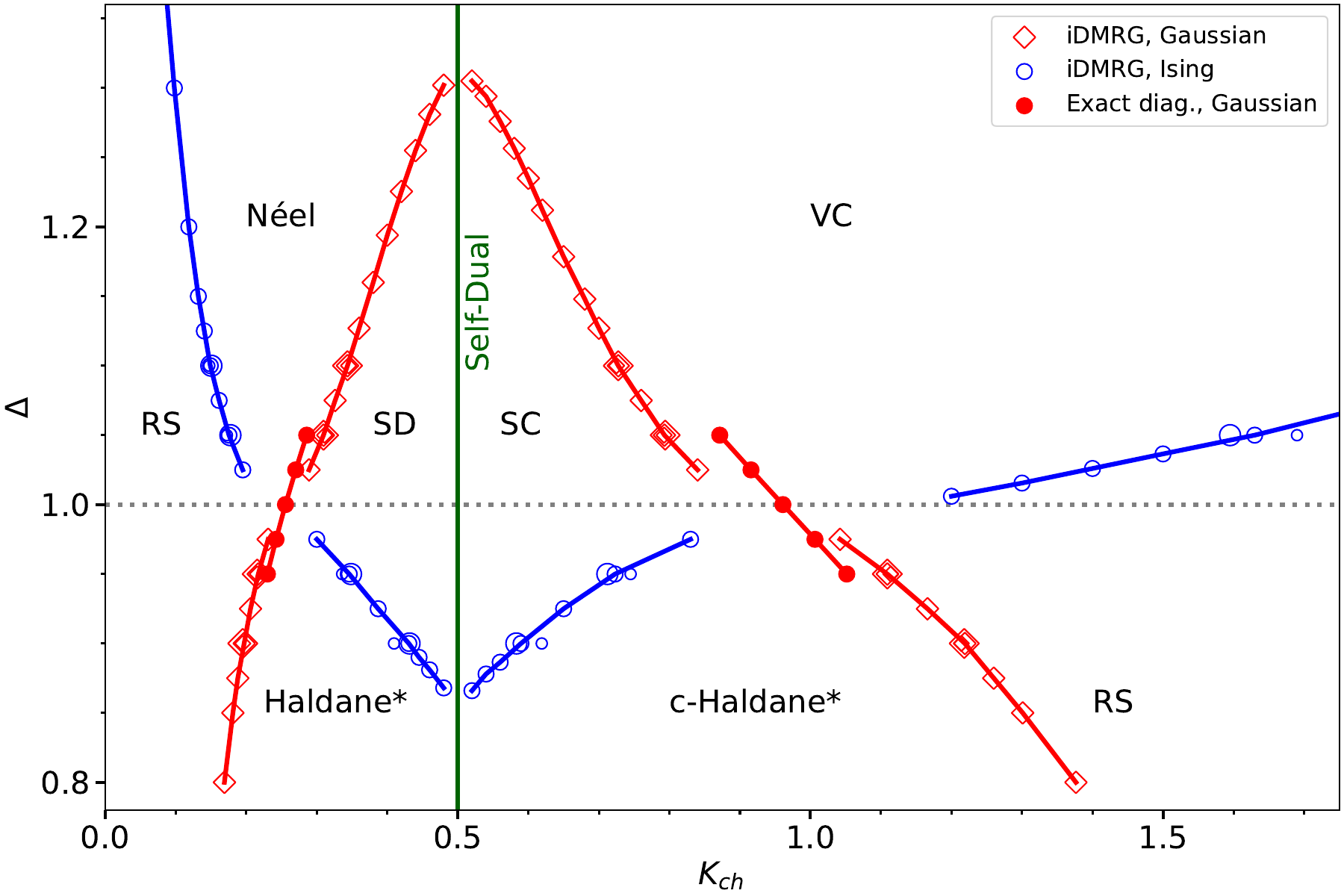}
\caption{\label{fig:pd_iDMRG}
Ground-state phase diagram of the XXZ-CCI ladder \eqref{eq:H_XXZ_CCI} with $J=1$ and $\Jperp=0.5$, obtained numerically. 
Red and blue empty symbols indicate Gaussian and Ising transition points, respectively, that are determined by iDMRG from the (local) maxima of the correlation length. 
Medium-size empty symbols connected by solid lines show the data for the bond dimension $\chi=192$, 
which give semiquantitative phase boundaries. 
In some cases, the data for $\chi = 128$ and $\chi = 256$ are also presented with small and large empty symbols, respectively, 
to illustrate the $\chi$-dependence of the results. 
Near the isotropic case $\Delta=1$, where iDMRG tends to converge worse, 
we performed an exact diagonalization analysis of the polarization amplitude $z_{\rm rung}$, as shown in Fig.\ \ref{fig:J05_fit}. 
A sign change of $z_{\rm rung}$ corresponds to a Gaussian transition in the symmetric channel. 
Red filled symbols connected by solid lines show exact diagonalization results extrapolated to the infinite-size limit. 
}
\end{figure*}

In this section, we numerically investigate the ground-state phase diagram of the XXZ-CCI ladder \eqref{eq:H_XXZ_CCI} on the $K_{\rm ch}$-$\Delta$ plane, 
focusing on the case of $J=1$ and $\Jperp=0.5$. 
Namely, we start from the $\Jperp=0.5$ cut of Fig.\ \ref{fig:phases_ch_ed}, and extend the phase diagram by introducing the anisotropy $\Delta$. 
The obtained phase diagram is presented in Fig.\ \ref{fig:pd_iDMRG}. 
In agreement with Fig.\ \ref{fig:phases_chex}(b,d) predicted by the Abelian bosonization and the duality, 
the RS-SD and SC-RS transition points at $\Delta=1$ turned out to be crossing points of separate transition lines in the anisotropic case. 
Furthermore, in agreement with the hard-core bosons picture in Sec.\ \ref{sec:bos_lat}, 
direct N\'eel-VC and Haldane*-c-Haldane* transitions occur when $\Delta$ is sufficiently away from unity in the easy-axis and easy-plane regimes, respectively. 

To obtain the phase diagram in Fig.\ \ref{fig:pd_iDMRG}, we employed iDMRG as well as exact diagonalization. 
When the system is not too close to the isotropic case $\Delta=1$, 
transition points were estimated from peak positions in the correlation length calculated by iDMRG; 
details of this analysis will be explained later. 
The transition points estimated in this way are shown with empty symbols in Fig.\ \ref{fig:pd_iDMRG}. 
Near the isotropic case $\Delta=1$, however, we encountered a convergence issue in iDMRG. 
Specifically, in iDMRG calculations with a finite bond dimension $\chi$, 
we found some ranges near the RS-SD and SC-RS transitions over which a spontaneous N\'eel or VC order appeared even in the isotropic case $\Delta=1$. 
These ranges shrunk with an increase in $\chi$. 
Similar behavior was also discussed in Appendix B of Ref.\ \cite{Ogino21nvbs}. 
As spontaneous N\'eel and VC orders are prohibited in the ground state of 1D SU$(2)$-symmetric systems (unless the uniform magnetic susceptibility diverges) \cite{Momoi96}, 
their appearance in iDMRG must be an artifact due to finite $\chi$. 
Because of this convergence issue, transitions near $\Delta=1$ could not be detected via sharp peaks in the correlation length calculated by iDMRG. 
To determine transition points near $\Delta=1$, we instead analyzed the polarization amplitude $z_{\rm rung}$ by exact diagonalization, as we did in Fig.\ \ref{fig:J05_fit}. 
This analysis can detect the Gaussian transition in the symmetric channel, as seen in the bosonized expression \eqref{eq:z_rung_phi}. 
The transition points extrapolated to the infinite-size limit are shown by red filled circles in Fig.\ \ref{fig:pd_iDMRG}. 
The iDMRG and exact diagonalization results are essentially consistent in the regimes where transition points were calculated with both the methods. 
We infer that small differences between the results of the two methods are mainly due to 
the need of larger system sizes in exact diagonalization to properly perform infinite-size extrapolation. 
As the exact diagonalization analysis was done in the same way as the isotropic case, 
we focus on the iDMRG analysis in the following. 

\subsection{iDMRG calculations}\label{sec:iDMRG}

\newcommand{\Nuc}{N_{\rm uc}}

We performed iDMRG calculations with the C++ library ITensor \cite{itensor,itensor-r0.3}. 
In the iDMRG algorithm \cite{McCulloch07,McCulloch08}, one represents a many-body wave function in the infinite one-dimensional (1D) system 
by a periodic matrix product state (MPS) with a bond dimension $\chi$. 
In applying iDMRG to the present ladder geometry, we regard the two-leg ladder as a zigzag chain arranged in the following order: 
$\dots,\Sv_{1,j},\Sv_{2,j},\Sv_{1,j+1},\Sv_{2,j+1},\dots$. 
With this arrangement, we performed iDMRG using a MPS with a four-site (i.e., two-rung) period in most of the calculations. 
In this setup, the variational MPS can exhibit spontaneous ordering as in Fig.\ \ref{fig:laddermodel}(e,f,g,h), 
and the corresponding order parameter can be computed directly. 
Only in the calculations of topological indices in Sec.\ \ref{sec:topo}, we employed a MPS with a two-site (i.e., one-rung) period 
as the translational invariance along the leg is important in such calculations. 

In general, a continuous phase transition is associated with a divergence in the correlation length $\xi$. 
In iDMRG calculations, a genuine divergence of $\xi$ does not occur, and it is instead detected through a peak in $\xi$ that grows with an increase $\chi$. 
From the MPS, the correlation length $\xi$ can be calculated as 
\begin{equation}\label{eq:corr2}
 \xi(\chi) = \frac{-2}{\ln |\epsilon_2(\chi)| },
\end{equation}
where $\epsilon_2(\chi)$ is the second largest eigenvalue of the transfer matrix. 
Here, a factor of $2$ is multiplied as the transfer matrix is defined over two rungs. 
The convergence criterion for our calculation was that the relative change of the correlation length during the last $10$ sweeps,  
i.e., $(\xi_{\rm n_{\rm s}+10}-\xi_{\rm n_{\rm s}})/\xi_{\rm n_{\rm s}}$ with $n_{\rm s}$ being the number of sweeps, became less than $10^{-6}$. 

In Fig.\ \ref{fig:pd_iDMRG}, empty symbols show transition points that are determined by iDMRG from the (local) maxima of $\xi$. 
For $\Delta \in \{0.9,0.95,1.05,1.10\}$, in particular, transitions were investigated using three different bond dimensions ($\chi=128, 192, 256$). 
We find that the obtained transition points are slightly $\chi$-dependent for the Ising transitions and almost $\chi$-independent for the Gaussian transitions. 
For other values of $\Delta$, the bond dimension was fixed to $\chi=192$, 
which was sufficient for a semiquantative phase diagram; see medium-size empty symbols interpolated by solid lines. 

Near the transitions, we encountered some convergence issues. For each transition, we used the ground state 
calculated at a point far from the transition as the initial state for the calculation close to the transition. 
In Figs.\ \ref{fig:SC_VC}-\ref{fig:RS_c-H} presented below, data points calculated from a point on the left (right) side of the transition are plotted in red (blue); 
for each value of $K_{\rm ch}$ and $\chi$, we kept only the point with the lowest energy.

The phase diagram in Fig.\ \ref{fig:pd_iDMRG} shows two crosses of transition lines, 
in consistency with the field-theoretical prediction in Sec.\ \ref{sec:EFT}. 
The first cross appearing on the left side ($K_{\rm ch}<1/2$) was already investigated in detail 
in a similar model with an inter-leg DDI \cite{Ogino21nvbs, Ogino21spt}. 
The structure of the phase diagram and the natures of the transitions being the same, we will not discuss this regime further. 
The remainder of Sec.\ \ref{sec:numerics_anisotropic} will thus be devoted to the study of the right side ($K_{\rm ch}>1/2$) of the phase diagram.

\subsection{SC-VC transition}

\begin{figure}
\includegraphics[width=0.45\textwidth]{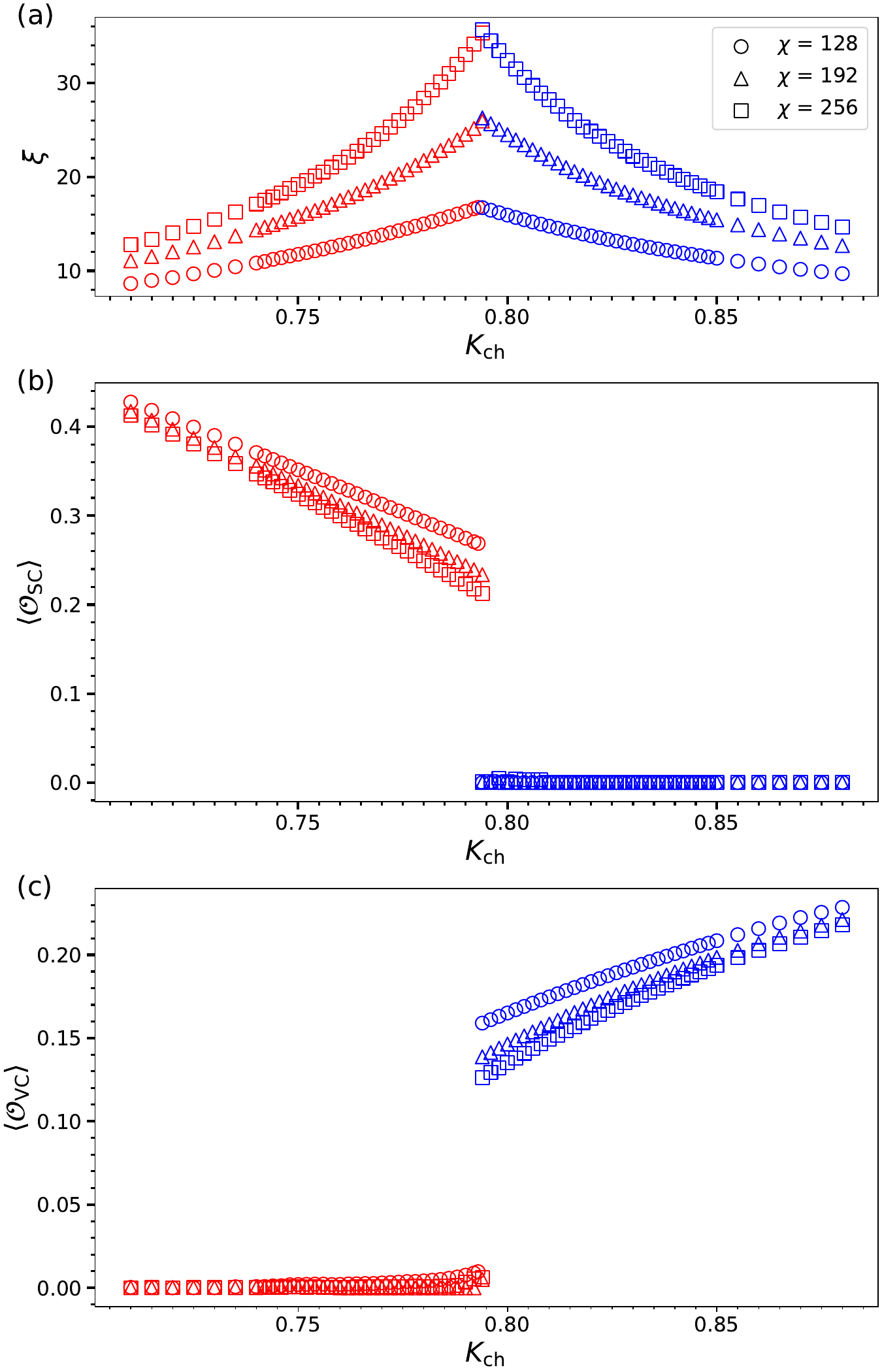}
\caption{\label{fig:SC_VC}
(a) Correlation length $\xi$ in Eq.\ \eqref{eq:corr2} and (b) the SC and VC order parameters in Eq.\ \eqref{eq:O_SC_VC}, 
calculated by iDMRG across the SC-VC Gaussian transition at $\Delta=1.05$ in Fig.\ \ref{fig:pd_iDMRG}. 
Red (blue) symbols indicate data points calculated using an initial point in the SC (VC) phase. 
Three different symbol shapes correspond to different $\chi$, as indicated in the legends. 
The maximum of $\xi$ in (a) is independent of $\chi$ within the parameter sampling of our calculation; 
we thus obtain a reasonable estimate $K_{\rm ch,c}=0.794(2)$ of the transition point. 
}
\end{figure}

\begin{figure}
\includegraphics[width=0.47\textwidth]{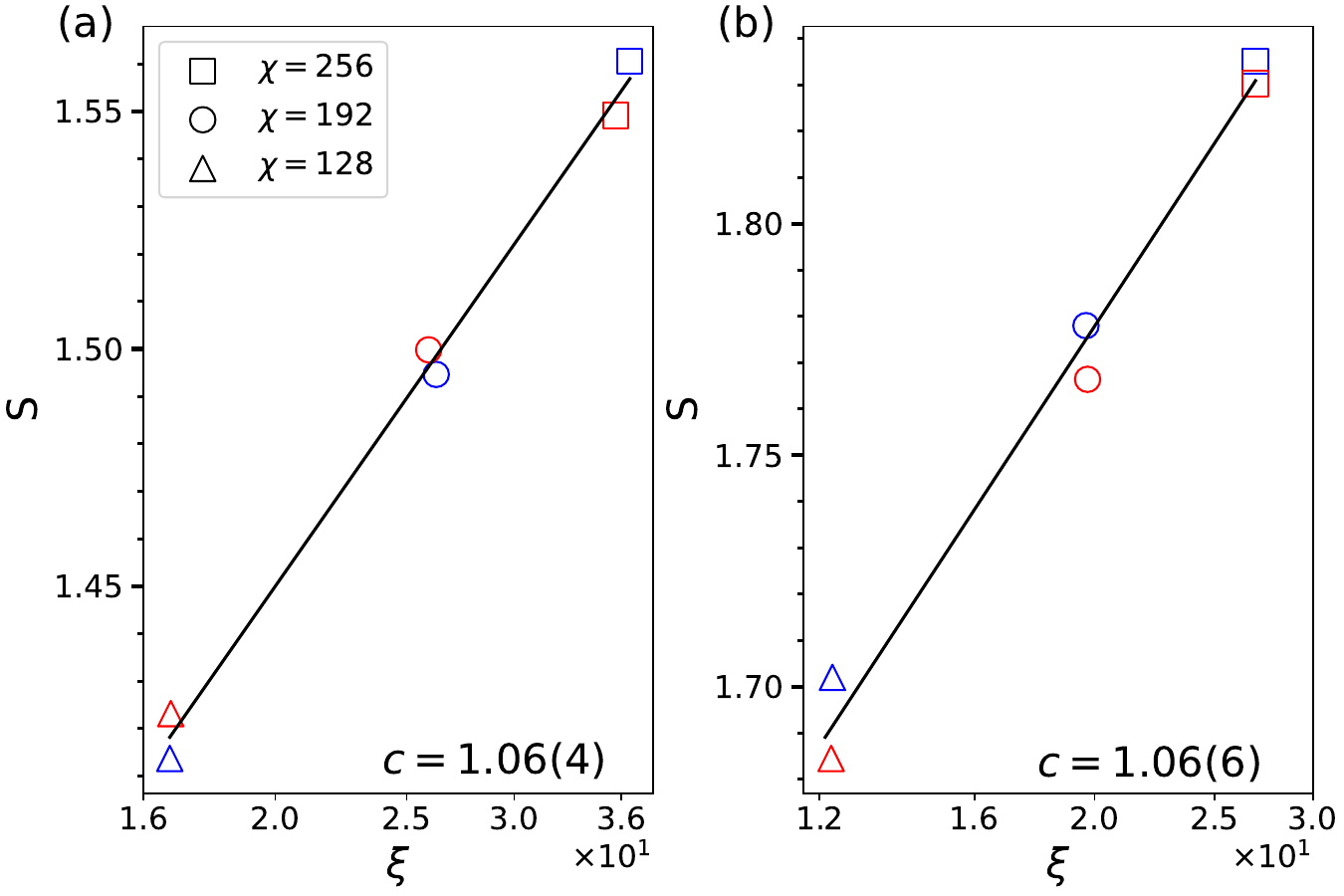}
\caption{\label{fig:CC_Gau}
Half-chain EE $S$ in Eq.\ \eqref{eq:EE} versus the correlation length $\xi$ in Eq.\ \eqref{eq:corr2}, 
calculated by iDMRG at (a) the SC-VC transition at $\Delta = 1.05$ in Fig.\ \ref{fig:SC_VC} 
and  (b) the c-Haldane*-RS transition at $\Delta = 0.95$ in Fig.\ \ref{fig:RS_c-H}. 
For each bond dimension $\chi$, the point with the highest correlation length in each phase is plotted. 
The same symbols and colors as in Figs.\ \ref{fig:SC_VC} and \ref{fig:RS_c-H} are used. 
A logarithmic scale is used for the horizontal axis. 
Solid lines show the fitting with Eq.\ \eqref{eq:CC}, 
from which the central charge $c$ is estimated as shown in each panel. 
Here, the number in parentheses in estimated $c$ shows an error in the last digit 
that is associated with the linear fitting. 
The results are in good agreement with the Gaussian universality class with $c=1$. 
}
\end{figure}

As discussed in Sec.\ \ref{sec:EFT}, the SC-VC transition is dual to the SD-N\'eel transition, and is expected to belong to the Gaussian universality class. 
This transition is an interesting example of a continuous transition between two ordered phases that break different symmetries. 
Based on its duality to the N\'eel-SD transition, the SC-VC transition gives a novel example of 
deconfined quantum critical points \cite{Senthil04} in one dimension \cite{JiangMotrunich19, Roberts19, Mudry19, Ogino21nvbs}. 

In Fig.\ \ref{fig:SC_VC}, we analyze the SC-VC transition at $\Delta=1.05$ as a representative case. 
As seen in Fig.\ \ref{fig:SC_VC}(a), the correlation length $\xi$ shows a maximum 
that grows with an increase in $\chi$ and whose position depends little on $\chi$. 
This growing maximum is indicative of a continuous phase transition. 
Furthermore, the little dependence of the peak position on $\chi$ allows us to obtain 
a reasonable estimate $K_{\rm ch,c}=0.794(2)$ of the transition point without performing an extrapolation as a function of $\chi$. 
Here, the number in parentheses represents the standard error in the last digit(s). 
As shown in Fig.\ \ref{fig:SC_VC}(b), 
each phase is characterized by nonzero values of the corresponding (SC or VC) order parameter. 
As one approaches the estimated transition point, 
the order parameter decreases continuously, and suddenly drops to zero at the transition. 
It is difficult to find a continuous change of the order parameter 
($\expval{\mathcal{O}_{\text{SC,VC}}} \sim |K_{\rm ch}-K_{\rm ch,c}|^\beta$ with $\beta=K_+/(4-4K_+)$ dependent on the TLL parameter $K_+$) 
across the transition in a simulation with finite $\chi$; 
to do so, an appropriate extrapolation of the order parameters as a function of $\chi$ will be necessary, which we do not address in the present paper. 
We refer the reader to Ref.\ \cite{Ogino21nvbs}, 
where critical behavior of the correlation length and order parameters were analyzed in detail for the N\'eel-SD transition, 
i.e., the dual counterpart of the SC-VC transition investigated here. 

To investigate the universality class of the transition, it is useful to examine the entanglement entropy (EE). 
For this purpose, one often calculates in iDMRG the half-chain EE 
\begin{equation}\label{eq:EE}
S = -\sum_{\alpha=1}^\chi \lambda_\alpha^2 \ln \lambda_\alpha^2, 
\end{equation}
where $\{\lambda_\alpha\}$ is a set of Schmidt coefficients associated with the bipartition of the infinite system into the left and right halves. 
According to conformal field theory, the half-chain EE $S$ and the correlation length $\xi$ have the relationship \cite{Calabrese04,Pollmann09}
\begin{equation}\label{eq:CC}
 S = \frac{c}{6} \ln \xi + S_0,
\end{equation}
where $c$ is the central charge and $S_0$ is a non-universal constant. 
In Fig.\ \ref{fig:CC_Gau}(a), we plot the relation between $S$ and $\xi$ calculated by iDMRG at the SC-VC transition. 
The data well obey the scaling form \eqref{eq:CC}, 
and the central charge $c=1.06(4)$ 
obtained from the fitting is in good agreement with the Gaussian universality class with $c=1$. 

\subsection{VC-RS transition}

\begin{figure}
\includegraphics[width=0.47\textwidth]{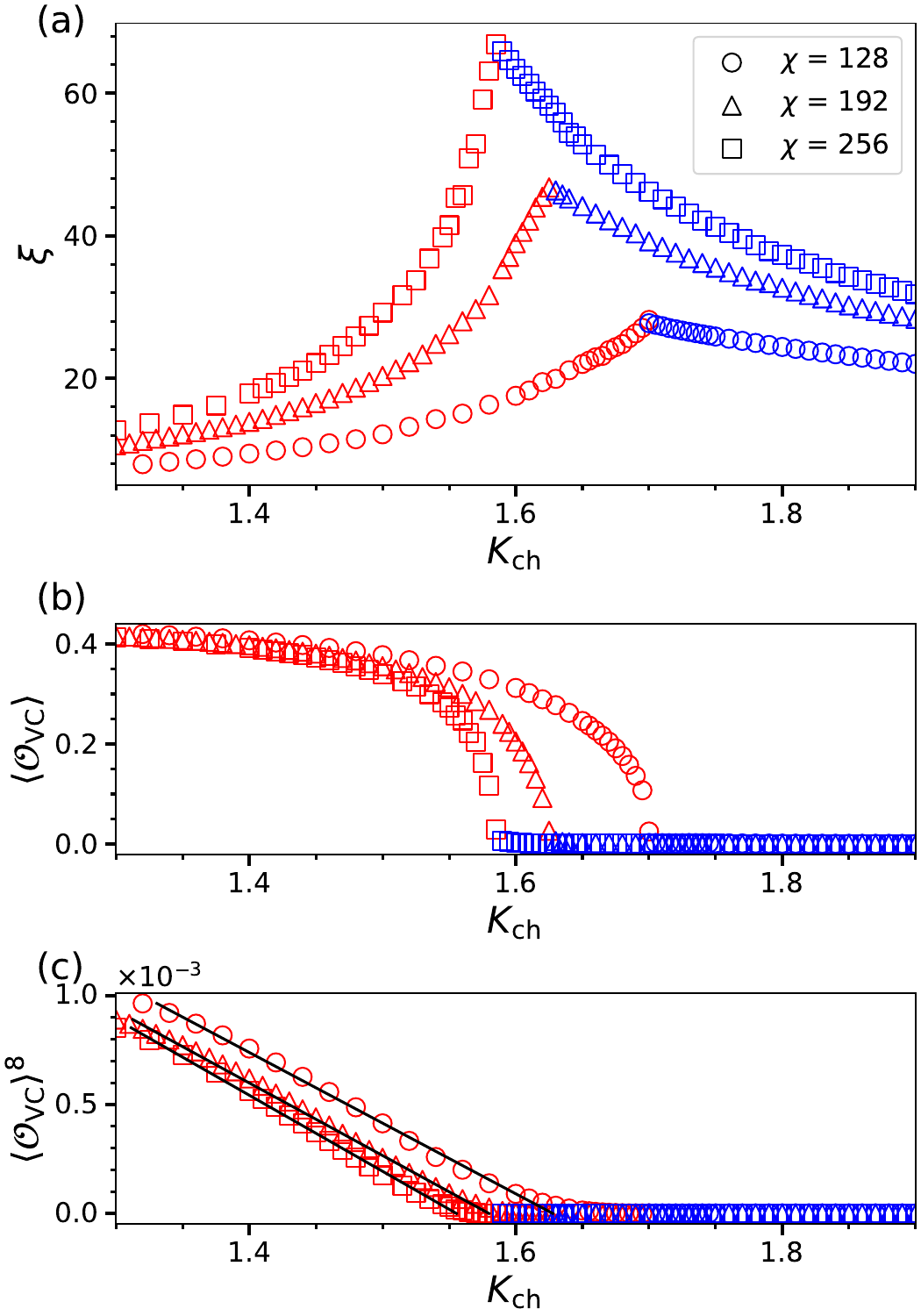}
\caption{\label{fig:VC_RS_cor_odp}
(a) Correlation length $\xi$ in Eq.\ \eqref{eq:corr2}, 
(b) the VC order parameter $\langle\mathcal{O}_{\rm{VC}}\rangle$ in Eq.\ \eqref{eq:O_VC}, and (c) $\langle \mathcal{O}_{\rm{VC}}\rangle^8$, 
calculated by iDMRG across the VC-RS transtion at $\Delta = 1.05$ in Fig.\ \ref{fig:pd_iDMRG}. 
Red (blue) symbols indicate data points calculated using an initial point in the VC (RS) phase. 
In (a), a maximum of $\xi$ occurs at $K_{\rm ch,c}\approx 1.70$, $1.63$, and $1.59$ for $\chi = 128$, $192$, and $256$, respectively. 
In (c), a linear fitting is done for each $\chi$ in the regime not to close to the transition; see solid lines. 
The intercept with the horizontal axis gives $K_{\rm ch,c}\approx 1.63$, $1.58$, and $1.55$ for $\chi = 128$, $192$, and $256$, respectively. 
The difference between the two estimates of the transition point becomes smaller with an increase in $\chi$. 
%
}
\end{figure}

\begin{figure}
\includegraphics[width=0.47\textwidth]{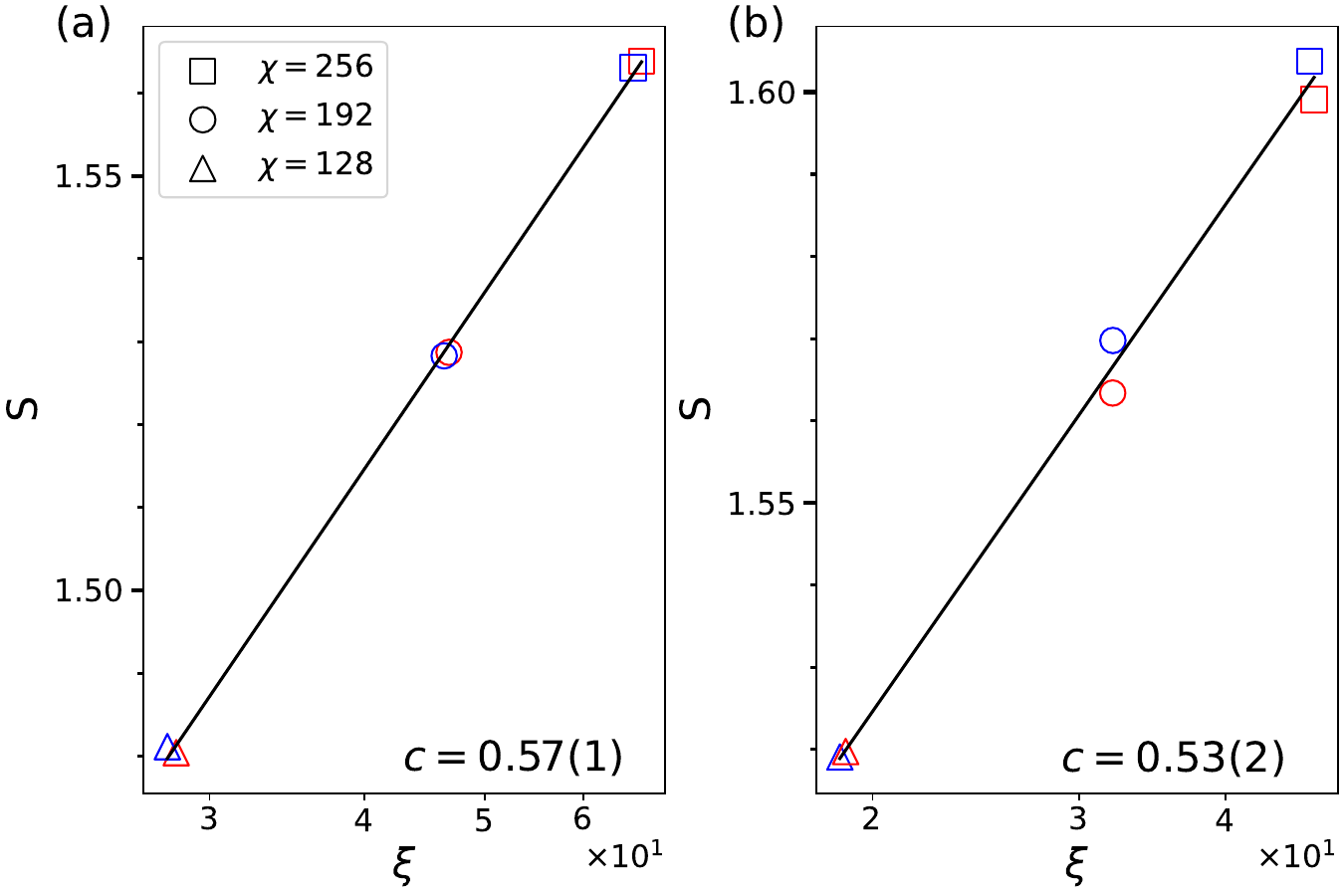}
\caption{\label{fig:CC_Ising}
Half-chain EE $S$ in Eq.\ \eqref{eq:EE} versus the correlation length $\xi$ in Eq.\ \eqref{eq:corr2}, calculated by iDMRG 
at (a) the VC-RS transition at $\Delta = 1.05$ in Fig.\ \ref{fig:VC_RS_cor_odp}
and (b) SC-c-Haldane* transition at $\Delta = 0.95$ in Fig.\ \ref{fig:SC_c-H_cor_odp}. 
For each bond dimension $\chi$, the point with the highest correlation length in each phase is plotted. 
The same symbols and colors as in Figs.\ \ref{fig:VC_RS_cor_odp} and \ref{fig:SC_c-H_cor_odp} are used. 
A logarithmic scale is used for the horizontal axis. 
Solid lines show the fitting with Eq.\ \eqref{eq:CC}; 
the estimated central charge $c$ shown in each panel agrees reasonably with the Ising universality class with $c=1/2$. 
}
\end{figure}

The VC-RS transition is dual to the N\'eel-RS transition, and is expected to belong to the Ising universality class. 
In Fig.\ \ref{fig:VC_RS_cor_odp}, we analyze the VC-RS transition at $\Delta=1.05$ as a representative case. 
As seen in Fig.\ \ref{fig:VC_RS_cor_odp}(a), the maximum of $\xi$ grows with an increase in $\chi$, 
which is a sign of a continuous transition. 
This maximum of $\xi$ is associated with the onset of the VC order, 
as seen in the plot of the VC order parameter $\langle\mathcal{O}_{\rm{VC}}\rangle$ in Fig.\ \ref{fig:VC_RS_cor_odp}(b). 
Meanwhile, the position of the maximum of $\xi$ moves gradually to the left with an increase in $\chi$. 
This indicates that a certain extrapolation to the infinite-$\chi$ limit is necessary to precisely determine the transition point. 
In the absence of a systematic extrapolation method, in the phase diagram in Fig.\ \ref{fig:pd_iDMRG}, 
we used the maximum of $\xi$ for $\chi=192$ to semiquantitatively draw the Ising transition lines (medium-size blue circles interpolated by solid lines). 

To investigate the universality class of the transition, we plot in Fig.\ \ref{fig:CC_Ising}(a) 
the relation between the EE $S$ and the correlation length $\xi$ at the maxima in Fig.\ \ref{fig:VC_RS_cor_odp}(a). 
A fit with the scaling form \eqref{eq:CC} gives $c=0.57(1)$, 
which agrees reasonably with the Ising universality class with $c=1/2$. 

In the Ising universality class, the critical exponent for the order parameter is known to be $\beta=1/8$. 
The order parameter to the 8th power, $\langle\mathcal{O}_{\rm{VC}}\rangle^8$, is thus expected to behave linearly around the transition. 
In Fig.\ \ref{fig:VC_RS_cor_odp}(c), $\langle\mathcal{O}_{\rm{VC}}\rangle^8$ is indeed fit well with a straight line for each $\chi$. 
Some deviation from the line is seen near the transition, where finite-$\chi$ effects are more significant; 
however, this deviation becomes smaller with an increase in $\chi$, giving another support for the Ising universality class. 
The intercept of the line with the horizontal axis gives an alternative estimate of the transition point. 
The transition points estimated in this way show a smaller $\chi$-dependence than those from the maximum of $\xi$; 
see the estimates in the caption of Fig.\ \ref{fig:VC_RS_cor_odp}. 
However, we note that a linear fitting of $\langle\mathcal{O}_{\rm{VC}}\rangle^8$
involves some arbitrariness in the choice of the fitting region, 
which is particularly crucial in the regime where the order parameter does not grow up to large values. 
We therefore employed the simple estimates from the maximum of $\xi$ in the phase diagram in Fig.\ \ref{fig:pd_iDMRG}.

 
\subsection{SC-c-Haldane* transition}

\begin{figure}
\includegraphics[width=0.45\textwidth]{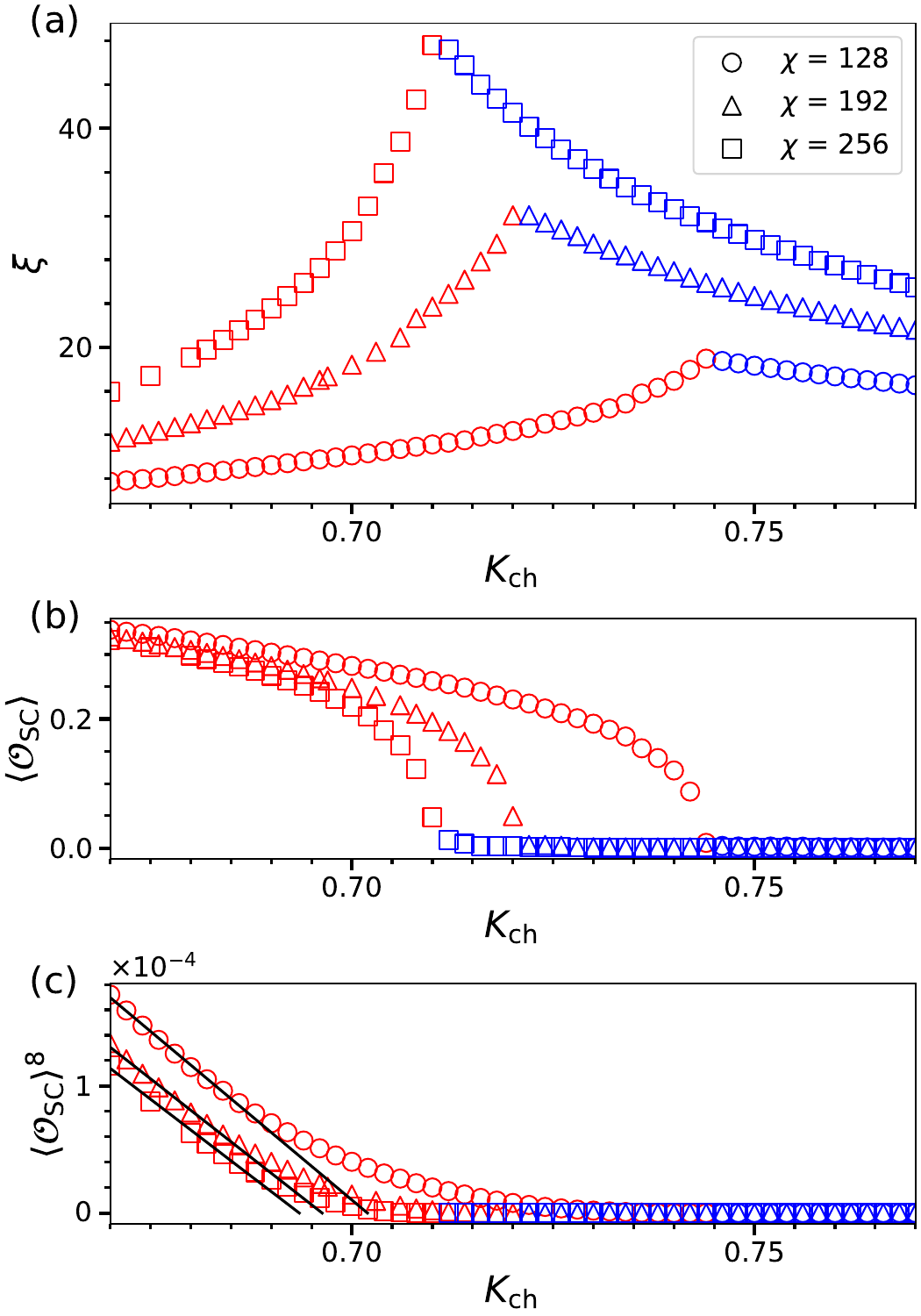}
\caption{\label{fig:SC_c-H_cor_odp}
(a) Correlation length $\xi$ in Eq.\ \eqref{eq:corr2}, 
(b) the SC order parameter $\langle\mathcal{O}_{\rm{SC}}\rangle$ in Eq.\ \eqref{eq:O_SC}, and (c) $\langle \mathcal{O}_{\rm{SC}}\rangle^8$, 
calculated by iDMRG across the SC-c-Haldane* transtion at $\Delta = 0.95$ in Fig.\ \ref{fig:pd_iDMRG}. 
Red (blue) symbols indicate data points calculated using an initial point in the SC (c-Haldane*) phase. 
In (a), a maximum of $\xi$ occurs at $K_{\rm ch}\approx 0.75$, $0.72$, and $0.71$ for $\chi = 128$, $192$, and $256$, respectively. 
In (c), a linear fitting is done for each $\chi$, as shown by solid lines. 
The intercept with the horizontal axis gives $K_{\rm ch}\approx 0.701$, $0.696$, and $0.693$ for $\chi = 128$, $192$, and $256$, respectively. 
%
}
\end{figure}

The SC-c-Haldane* transition is dual to the SD-Haldane* transition, and is also expected to belong to the Ising universality class. 
In Fig.\ \ref{fig:SC_c-H_cor_odp}, we analyze the SC-c-Haldane* transition at $\Delta=0.95$ as a representative case. 
As seen in Fig.\ \ref{fig:SC_c-H_cor_odp}(a,b), the correlation length $\xi$ shows a maximum, 
which is associated with the onset of the SC order. 
The position of the maximum again moves gradually to the left with an increase in $\chi$, 
indicating a difficulty in precisely determining the transition point in finite-$\chi$ simulations. 
In Fig.\ \ref{fig:CC_Ising}(b), we plot the relation between $S$ and $\xi$ at the maxima in Fig.\ \ref{fig:SC_c-H_cor_odp}(a); 
the slope of the $S$-$\xi$ relation agrees well with the Ising universality class with $c=1/2$.  
In Fig.\ \ref{fig:SC_c-H_cor_odp}(c), we estimate the transition point 
by fitting $\langle \mathcal{O}_{\rm{SC}}\rangle^8$ by a line and finding the intercept with the horizontal axis. 
The $\chi$-dependence of the transition point estimated in this way is relatively small; see the estimates in the caption of Fig.\ \ref{fig:SC_c-H_cor_odp}. 
However, deviations from the lines seem more significant than the case of Fig.\ \ref{fig:VC_RS_cor_odp}(c), 
owing to the presence of another transition at the self-dual line $K_{\rm ch}=1/2$ that prevents the order parameter to fully increase. 
Therefore, there is larger ambiguity in the choice of the fitting region. 

\subsection{Topological transitions between featureless phases}\label{sec:topo}

\begin{figure}
\includegraphics[width=0.45\textwidth]{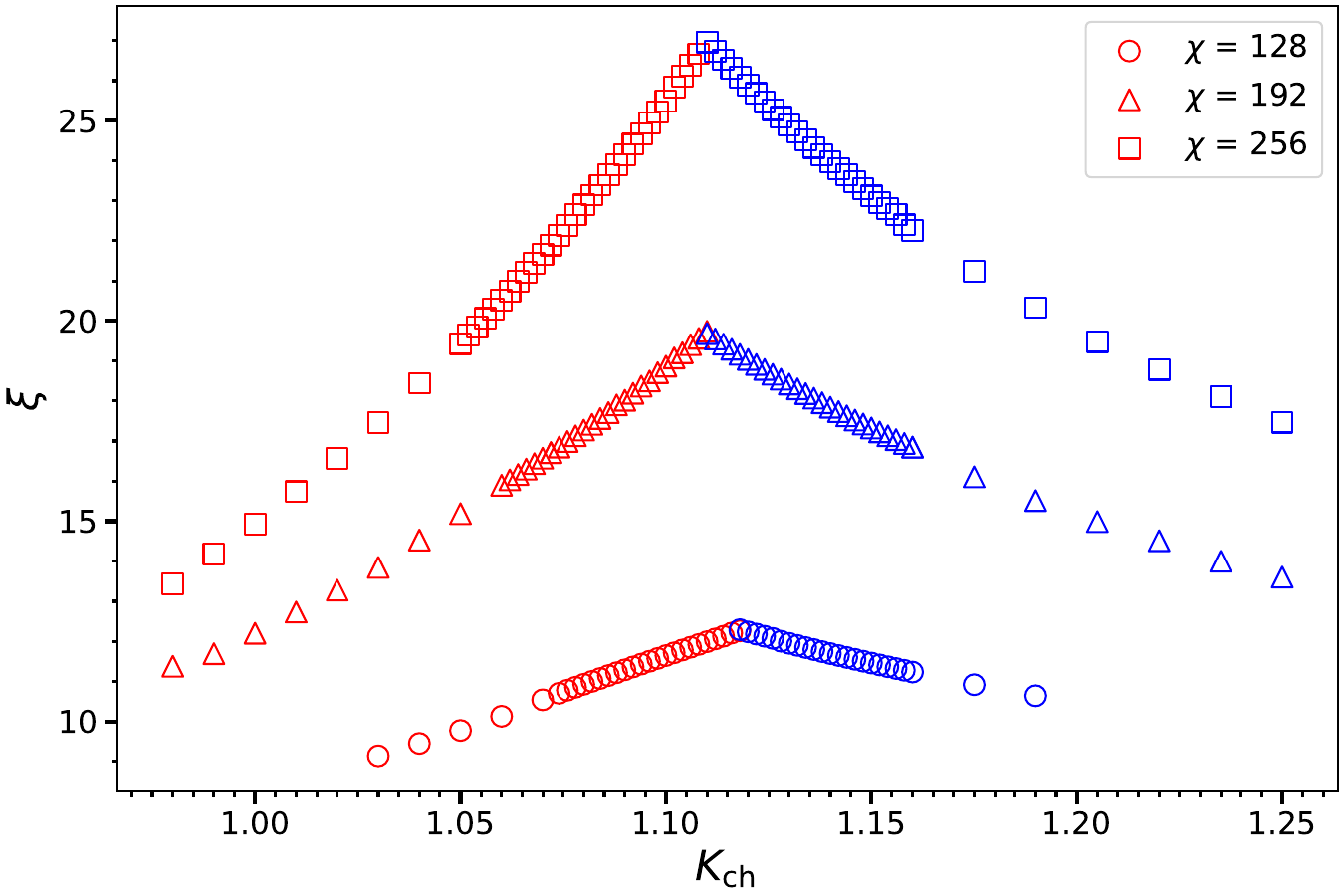}
\caption{\label{fig:RS_c-H}
Correlation length $\xi$ in Eq.\ \eqref{eq:corr2}, calculated by iDMRG across the c-Haldane*-RS transition at $\Delta = 0.95$ in Fig.\ \ref{fig:pd_iDMRG}. 
Red (blue) symbols indicate data points calculated using an initial point in the c-Haldane* (RS) phase. 
The peak position of $\xi$ depends little on $\chi$; 
in particular,  the peak positions for $\chi=192$ and $256$ are precisely the same within the parameter sampling of our calculation, 
giving a reasonable estimate $K_{\rm ch,c}=1.110(2)$ of the transition point. 
}
\end{figure}

The c-Haldane*-RS transition is dual to the Haldane*-RS transition, and is expected to belong to the Gaussian universality class. 
In Fig.\ \ref{fig:RS_c-H}, we examine the correlation length $\xi$ across the c-Haldane*-RS transition at $\Delta = 0.95$. 
Again, the growing maximum of $\xi$ is indicative of a continuous transition. 
Furthermore, the little dependence of the peak position of $\xi$ allows us to obtain a reasonable estimate $K_{\rm ch,c}=1.110(2)$ of the transition point. 
The central charge of the transition is examined by plotting the $S$-$\xi$ relation in Fig.\ \ref{fig:CC_Gau}(b), 
showing a good agreement with the Gaussian universality class with $c=1$. 

The c-Haldane*-RS transition is an example of a topological phase transition---it  
occurs between two featureless phases and is not associated with a spontaneous symmetry breaking. 
Therefore, this transition cannot be characterized by the onset of a certain local order parameter. 
Similarly, the RS-Haldane* and Haldane*-c-Haldane* transitions are also examples of topological phase transitions. 
As shown by Liu {\it et al.}\ \cite{LiuZX12}, the RS, Haldane, and Haldane* phases are distinct phases in the presence of the $D_2 \times \sigma$ symmetry, 
where $D_2=\mathbb{Z}_2\times\mathbb{Z}_2$ is the discrete spin rotational symmetry and $\sigma$ is the leg-interchange symmetry. 
These phases can thus be distinguished by topological indices \cite{Pollmann10, Pollmann12, PollmannTurner12, ChenX11, ChenX11_2,Schuch11, Tasaki20book}
associated with $D_2 \times \sigma$, 
as numerically demonstrated by Ogino {\it et al.}\ \cite{Ogino21spt,Ogino22} in related ladder models. 
However, we now have an additional phase, i.e., the c-Haldane* phase. 
This phase is not distinguished from the Haldane* phase by the topological indices associated with $D_2\times\sigma$; 
this is seen from the fact that the duality transformation $\Ucal({\pi/2})$ in Eq.\ \eqref{eq:U_theta} that relates these phases commutes with $D_2\times\sigma$. 
As we shall see, the Haldane* and c-Haldane* phases are distinguished by a topological index associated with time-reversal symmetry. 

\begin{figure}
\includegraphics[width=0.45\textwidth]{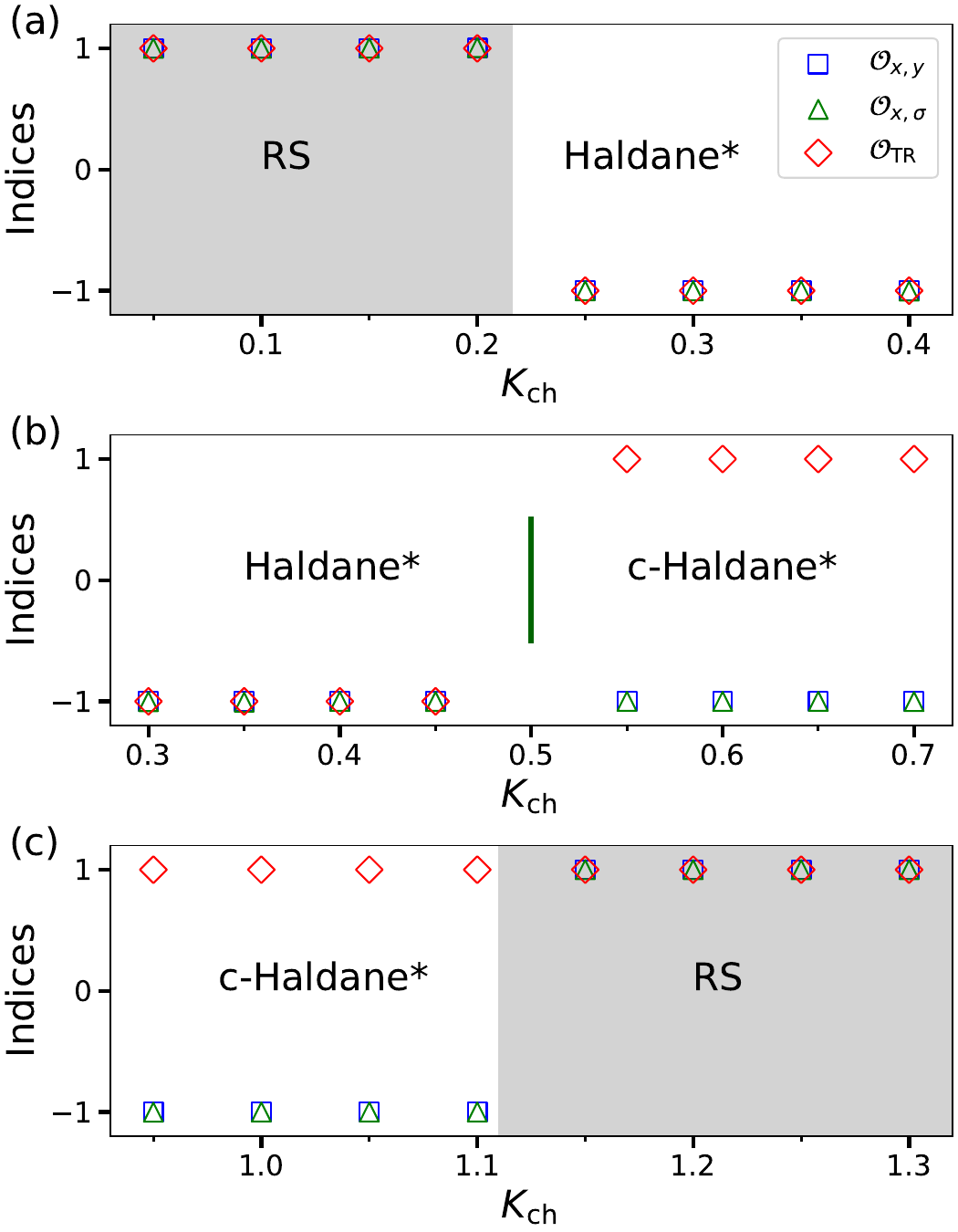}
\caption{\label{fig:topo}
Topological indices \eqref{eq:Oxy_Oxsig_Otr} calculated by iDMRG 
across (a) the RS-Haldane* transition at $\Delta=0.95$ 
(b) the Haldane*-c-Haldane* transition at $\Delta=0.7$, 
and (c) the c-Haldane*-Haldane transition at $\Delta=0.95$. 
The green vertical line in (b) indicates the self-dual point $K_{\rm ch}=1/2$. 
The boundaries between the shaded and unshaded regions in (a,c) correspond to the phase boundaries 
obtained from the maximum of $\xi$; see, e.g., Fig.\ \ref{fig:RS_c-H}. 
}
\end{figure}

Let us briefly review how to calculate the topological indices in the MPS representation. 
For the $D_2 \times \sigma$ symmetry, more detailed information is provided in Ref.\ \cite{Ogino21spt}; 
here, we also discuss the index associated with time-reversal symmetry \cite{Pollmann10, Tasaki20book}.  
Assuming the translational invariance, the ground state of the infinite system can be represented in the form of a canonical MPS 
\begin{equation}\label{eq:uMPS}
\ket{\Psi} = \sum_{\dots,l,m,n,\dots} 
[ \dots \Lambda \Gamma_{l} \Lambda \Gamma_{m} \Lambda \Gamma_{n} \dots] \ket{\dots l,m,n, \dots}, 
\end{equation}
where $\Gamma_m$ is a $\chi\times\chi$ matrix with $m$ being the spin state on a ``site'',
and $\Lambda=\mathrm{diag}(\lambda_1,\dots,\lambda_\chi)$ is a diagonal matrix comprised of Schmidt coefficients. 
When applied to the spin-$\frac12$ ladder, a ``site'' corresponds to a rung, and $m$ runs over the four spin states on a rung. 
In our iDMRG calculations, we employed a periodic MPS representation along the zigzag chain as explained in Sec.\ \ref{sec:iDMRG}; 
therefore, the matrix $\Gamma_m$ in Eq.\ \eqref{eq:uMPS} is obtained by combining two matrices on a rung. 

Suppose that $\ket{\Psi}$ is invariant under an on-site unitary transformation, 
which is represented as a unitary matrix $\Sigma_{mm'}$ acting on the spin indices on a ``site''. 
Then the $\Gamma_m$ matrices can be shown to satisfy \cite{PerezGarcia08}  
\begin{align}\label{eq:defU}
 \sum_{m'} \Sigma_{mm'} \Gamma_{m'} = e^{i\theta_\Sigma} U_\Sigma^\dagger \Gamma_m U_\Sigma, 
\end{align}
where $e^{i\theta_\Sigma}$ is a phase factor and $U_\Sigma$ is a $\chi\times\chi$ unitary matrix. 
Physically, $U_\Sigma$ describes a symmetry transformation acting on fictitious edge states in the entanglement Hamiltonian. 
For a symmetry group $\{\Sigma\}$, the phase factors $\{ e^{i\theta_\Sigma} \}$ form a 1D representation of the group 
while the unitary matrices $\{ U_\Sigma \}$ form a projective representation of the group. 

Specifically, the $D_2\times\sigma$ symmetry of a spin ladder is generated by the following on-site transformations: 
the spin rotations $\Sigma^x := \exp[i\pi (S^x_1 + S^x_2)]$ and $\Sigma^y := \exp[i\pi (S^y_1 + S^y_2)]$ 
and the leg interchange $\Sigma^\sigma:=\sum_{\alpha,\beta=\uparrow,\downarrow}\ket{\alpha\beta}\bra{\beta\alpha}$. 
For these transformations, we can introduce the unitary matrices 
$U_x:=U_{\Sigma^x}$, $U_y:=U_{\Sigma^y}$, and $U_\sigma:=U_{\Sigma^\sigma}$ via Eq.\ \eqref{eq:defU}. 
While $\Sigma^x$, $\Sigma^y$, and $\Sigma^\sigma$ commute with one another, 
the commutation relations among $U_x$, $U_y$, and $U_\sigma$ may involve nontrivial phase factors, 
which are signs of topologically nontrivial phases. 
To illustrate this, the matrices $U_x$, $U_y$, and $U_\sigma$ (and also $U_{\rm TR}$ discussed below) are explicitly calculated 
for simple representative MPSs for the Haldane, Haldane*, and c-Haldane* phases in Appendix \ref{app:topo_MPS}; 
for example, the nontrivial relation $U_xU_y=-U_yU_x$ can be seen in all the three states in Table \ref{tab:U_SPT}. 

If $\ket{\Psi}$ is invariant under time reversal, a relation similar to Eq.\ \eqref{eq:defU} holds; 
however, one has to note that time reversal is defined by an on-site unitary transformation $\Sigma^y:=\exp[i\pi \left(S_1^y+S_2^y \right)]$ followed by complex conjugation. 
We therefore have 
\begin{align}\label{eq:defUtr}
 \sum_{m'} \Sigma^y_{mm'}\Gamma_{m'}^*= e^{i\theta_{\rm TR}} U_{\rm TR}^\dagger \Gamma_m U_{\rm TR},
\end{align}
where $e^{i\theta_{\rm TR}}$ is a phase factor and $U_{\rm TR}$ is a unitary matrix. 
One can show $U_{\rm TR} U_{\rm TR}^*=\mu I$ with $\mu=\pm 1$ \cite{Pollmann10, Tasaki20book}, where $I$ is the identity matrix. 
The case of $\mu=-1$ corresponds to a nontrivial phase with Kramers degeneracy in the entanglement spectrum. 

Numerically, the phase factor $e^{i\theta_\Sigma}$ and the matrix $U_\Sigma$ in Eq.\ \eqref{eq:defU} or \eqref{eq:defUtr} 
can be calculated from the eigenvalue problem of a generalized transfer matrix, as described in Refs.\ \cite{PollmannTurner12, PerezGarcia08, Ogino21spt}. 
A possible nontrivial algebra obeyed by $\{U_\Sigma \}$ can then be detected by calculating traced commutators \cite{PollmannTurner12, Avakian24, Sorensen24} 
\begin{subequations}\label{eq:Oxy_Oxsig_Otr}
\begin{align}
\mathcal{O}_{xy} &:= \frac{1}{\chi} \tr( U_x U_y U_x^\dagger U_y^\dagger ),\\
\mathcal{O}_{x\sigma} &:= \frac{1}{\chi} \tr( U_x U_\sigma U_x^\dagger U_\sigma^\dagger ), \\
\mathcal{O}_{\rm TR} &:= \frac{1}{\chi} \tr( U_{\rm TR} U_{\rm TR}^* ). 
\end{align}
\end{subequations}

Figure \ref{fig:topo} shows the indices \eqref{eq:Oxy_Oxsig_Otr} calculated by iDMRG for the three types of topological phase transitions in our XXZ-CCI ladder model. 
All these indices are constant in each phase, and take $\pm 1$ depending on the phase. 
All the four featureless phases are clearly distinguished by these indices. 
In particular, the Haldane* and c-Haldane* phases have the same indices in terms of the $D_2\times \sigma$ symmetry, 
but are distinguished by the index associated with time-reversal symmetry $\mathbb{Z}_2^T$. 

Historically, string order parameters have been used to characterize the Haldane phases of spin-$S$ chains with $S=\text{odd}$ \cite{denNijs89,KennedyTasaki92,Oshikawa92}. 
For a ladder problem, two types of string order parameters have been introduced, based on the total spins on rungs or diagonal lines
\cite{Watanabe93,Nishiyama95,Shelton96,Kim00,Kim08,Nakamura03}. 
For the present problem, these string order parameters can distinguish the RS phase from the Haldane, Haldane*, and c-Haldane* phases 
but cannot distinguish among the latter three; 
see, e.g., Ref.\ \cite{Ogino21spt} for a plot of them across the RS-Haldane* transition. 
Therefore, they give the same level of distinguishability as the index $\mathcal{O}_{xy}$. 
Our characterization based on the three indices \eqref{eq:Oxy_Oxsig_Otr} is thus more complete in analyzing gapped featureless phases 
under the $D_2\times\sigma\times \mathbb{Z}_2^T$ symmetry. 
As an alternative method, we refer the reader to Refs.\ \cite{PerezGarcia08,PollmannTurner12} 
for generalizations of nonlocal order parameters for other symmetry groups than $D_2=\mathbb{Z}_2\times\mathbb{Z}_2$. 
We also note that different generalizations of string order parameters have recently been proposed to characterize a variety of phases 
in extended Kitaev models on the ladder \cite{Aprapidis19,Catuneanu19,Sorensen24,Pandey22}.

\subsection{Criticality on the self-dual line}

\begin{figure}
\includegraphics[width=0.43\textwidth]{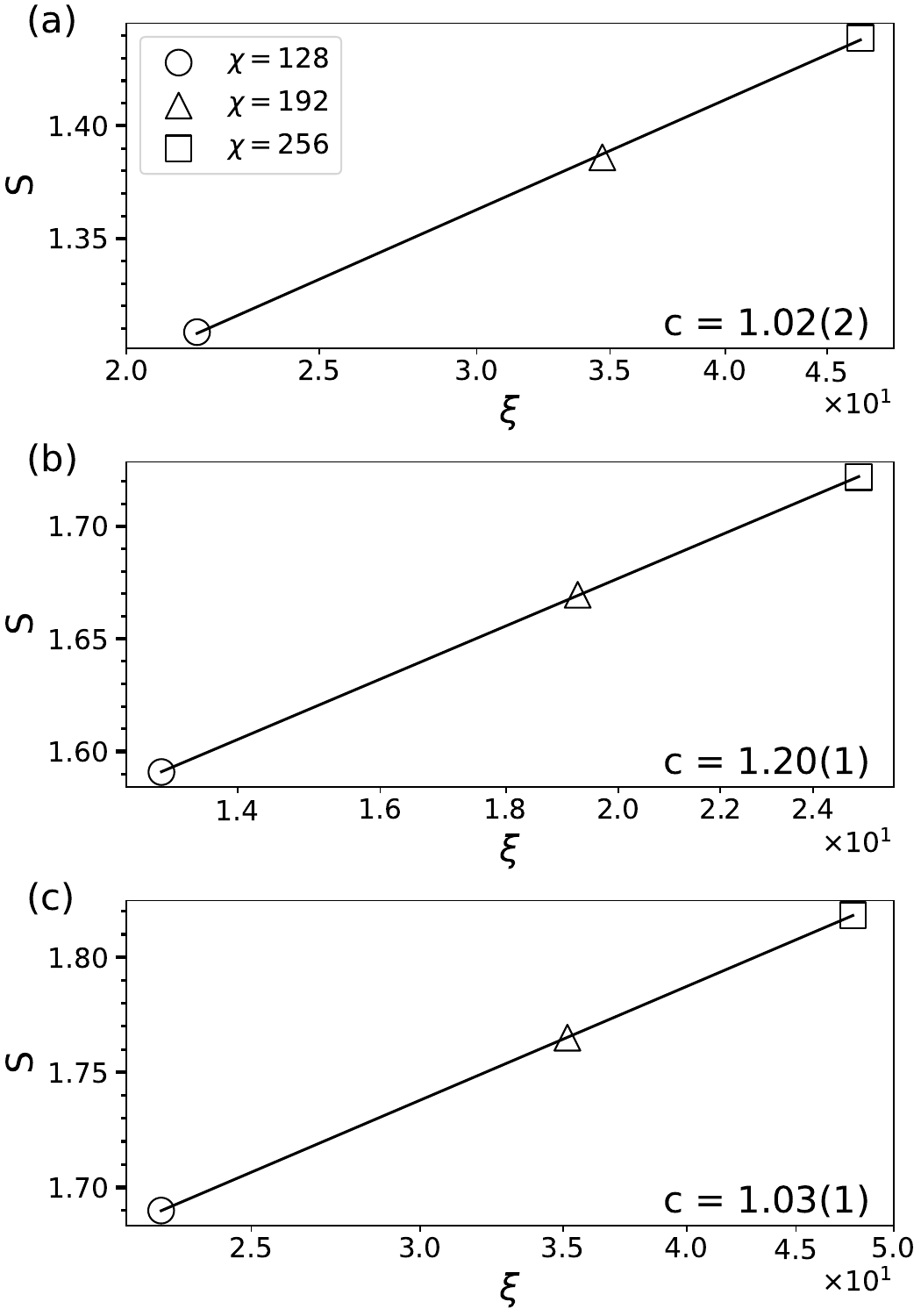}
\caption{\label{fig:CC_selfdual}
Half-chain EE $S$ in Eq.\ \eqref{eq:EE} versus the correlation length $\xi$ in Eq.\ \eqref{eq:corr2}, 
calculated by iDMRG on the self-dual line $K_{\rm ch}=1/2$ in Fig.\ \ref{fig:pd_iDMRG}. 
(a) The N\'eel-VC transition at $\Delta=1.8$. 
(b) The SD-SC transition at $\Delta=1$. 
(c) The Haldane*-c-Haldane* transition at $\Delta=0.6$. 
A logarithmic scale is used for the horizontal axis. 
Solid lines show the fitting with Eq.\ \eqref{eq:CC}, 
and the estimated central charge $c$ is shown in each panel. 
}
\end{figure}

Lastly, we investigate the central charge $c$ of the transitions on the self-dual line $K_{\rm ch}=1/2$ 
by examining the $S$-$\xi$ relation, as shown in Fig.\ \ref{fig:CC_selfdual}. 
The estimated values of $c$ for (a) the N\'eel-VC transition and (c) the Haldane*-c-Haldane* transition 
agree well with the Gaussian universality class with $c=1$ predicted in the hard-core bosons approach in Sec.\ \ref{sec:bos_lat}; see Fig.\ \ref{fig:phases_selfdual}. 
Meanwhile, for (b) the SD-SC transition at $\Delta=1$, we have not been able to confirm a good agreement 
with the $c=1$ Gaussian transition predicted by the field theory around the SU$(4)$ critical point \cite{Lecheminant05, Lecheminant06}. 
In view of the relatively small correlation length $\xi\approx 25$ even for largest $\chi$ in Fig.\ \ref{fig:CC_selfdual}(b), 
simulations with larger $\chi$ will be necessary to estimate $c$ more precisely. 

\section{Summary and outlook}\label{sec:summary}

In this paper, we have studied the ground-state phase diagram of the XXZ-CCI ladder \eqref{eq:H_XXZ_CCI} 
by means of the Abelian bosonization, the spin-chirality duality, spinor hard-core bosons approach, iDMRG, and exact diagonalization. 
We have obtained a rich phase diagram that consists of distinct gapped featulress and ordered phases, 
as shown in Figs.\ \ref{fig:phases_ch}, \ref{fig:phases_chex}, \ref{fig:phases_ch_ed}, and \ref{fig:pd_iDMRG}. 
In particular, we have obtained two crosses of transition lines, as shown in Fig.\ \ref{fig:pd_iDMRG}. 
The phase structure around the left cross, 
where the RS, Haldane*, SD and N\'eel phases appear, 
is similar to what has been found in a related ladder model with an inter-leg DDI \cite{Ogino21nvbs, Ogino21spt}. 
Our major contribution is on the phase structure around the right cross, 
where the RS, c-Haldane*, SC and VC phases appear as dual counterparts of the above phases. 
The Haldane* and c-Haldane* phases, which are dual to each other, can both be viewed as twisted variants of the Haldane phase on a ladder. 
However, a clear difference occurs in the appearance of complex elements in the c-Haldane state, 
which leads to a distinction under time-reversal symmetry. 
We have demonstrated that the RS, Haldane*, and c-Haldane* phases are distinguished by the topological indices \eqref{eq:Oxy_Oxsig_Otr}
associated with the $D_2\times\sigma$ symmetry and time-reversal symmetry $\mathbb{Z}_2^T$, as shown in Fig.\ \ref{fig:topo}. 
By means of the (spinor) hard-core bosons approach (Sec.\ \ref{sec:bos_lat}), 
we have analyzed critical behavior on the self-dual surface $K_{\rm ch}=J/2$. 
Specifically, we have argued that 
the N\'eel-VC transition in the easy-axis regime and the Haldane-c-Haldane* transition in the easy-plane regime 
both belong to the Gaussian universality class with $c=1$; see Figs.\ \ref{fig:phases_selfdual} and \ref{fig:phases_ch_Ising}. 
This is confirmed by estimating the central charge $c$ from the plot of the half-chain EE $S$ versus the correlation length $\xi$, as shown in Fig.\ \ref{fig:CC_selfdual}(a,c). 

An important future issue is to determine the phase diagram of 
a spin-$\frac12$ XXZ ladder with four-spin ring exchange, i.e., the XXZ-$K$ ladder in Eq.\ \eqref{eq:H_XXZ_K}. 
While the magnitudes of ring exchange $K$ are relatively small in cuprate ladders \cite{Brehmer99, Matsuda00}, 
possible effects of large $K$ have attracted much attention in the context of 
solid $^3$He \cite{Roger83,Fukuyama08, Misguich99} and Wigner crystals \cite{Chakravarty99, Klironomos07}. 
As argued in Sec.\ \ref{sec:XXZ_K}, the XXZ-$K$ ladder is expected to show an essentially similar phase diagram as the XXZ-CCI ladder studied in this paper. 
In a numerical study, however, special care must be taken to analyze the regime of large $K$, as noted in Sec.\ \ref{sec:XXZ_K}. 
A related intriguing issue is to study combined effects of XXZ anisotropy, frustration, and ring exchange 
by, e.g., extending the study of Metavitsiadis {\it et al.}\ \cite{Metavitsiadis17}. 
It will also be interesting to look for other examples of SPT phases 
in extended ladder systems with the $D_2\times\sigma\times Z_2^T$ symmetry,  
whose classification is given by the second group cohomology ${\cal H}^2 (\mathbb{Z}_2^3 \times \mathbb{Z}_2^T,U(1))=\mathbb{Z}_2^7$ \cite{ChenX11, ChenX11_2,Schuch11,YangLiuZX21,Ouyang21}. 

In the present work, the spin-chirality duality has played a key role in obtaining a rich variety of featureless and ordered phases. 
In this approach, transitions between mutually dual phases naturally occur on the self-dual surface. 
In particular, we have found an example of a continuous topological transition 
between the Haldane* and c-Haldane* phases that is protected by time-reversal symmetry. 
While the duality transformation $\Ucal(\pi/2)$ in Eq.\ \eqref{eq:U_theta} considered here is a local unitary transformation, 
nonlocal duality transformations have also been applied in studies of SPT phases and their transitions \cite{Verresen17,YangKatsura23,Chulliparambi23}. 
It will be interesting to apply duality more extensively 
as a guide to explore novel phases and transitions. 


\begin{acknowledgments}
The authors thank Takuhiro Ogino and Keisuke Totsuka for stimulating discussions. 
This research was supported by JSPS KAKENHI Grant Numbers JP18K03446, 
JP20H01849, JP21K03439, and JP23K03286 
and by the Center of Innovations for Sustainable Quantum AI (JST Grant No.\ JPMJPF2221). 
M.F. was also supported by Keio ``Design the Future'' Award for International Students. 
\end{acknowledgments}

\appendix
\section{Simple matrix product states under symmetry transformations}\label{app:topo_MPS}

\begin{table} 
\caption{\label{tab:U_SPT}
The phase factors $e^{i\theta_\Sigma}$ and the $2\times 2$ matrices $U_\Sigma$ in Eqs.\ \eqref{eq:defU_A} and \eqref{eq:defUtr_A}
for the spin rotations $\Sigma^x$ and $\Sigma^y$, the leg interchange $\Sigma^\sigma$, and time reversal 
in the representative Haldane, Haldane*, and c-Haldane* states described by Eqs.\ \eqref{eq:MPS_H}, \eqref{eq:MPS_Hs}, and \eqref{eq:MPS_cHs}. 
} 
\begin{ruledtabular} 
\begin{tabular}{c | cccccccc} 
State & $e^{i\theta_x}$ & $U_x$ & $e^{i\theta_y}$ & $U_y$ & $e^{i\theta_\sigma}$ & $U_\sigma$ &$e^{i\theta_{\rm TR}}$& $U_{\rm TR}$\\ 
\hline
H & $+1$ & $i\sigma^x$ &$+1$& $i\sigma^y$ &$+1$& $I$ &$+1$&$i\sigma^y$\\ 
H*& $-1$ & $i\sigma^x$ &$-1$& $i\sigma^y$ &$-1$& $i\sigma^z$ &$-1$&$i\sigma^y$\\ 
cH*& $-1$ &$i\sigma^x$ &$-1$& $i\sigma^y$ &$-1$& $i\sigma^z$ &$+1$&$i\sigma^x$
\end{tabular} 
\end{ruledtabular} 
\end{table}

In this appendix, we consider simple representative MPSs for the Haldane, Haldane*, and c-Haldane* phases of a spin-$\frac12$ ladder, 
and discuss their properties under symmetry transformations. 
Specifically, we will calculate the matrices $U_\Sigma$ in Eqs.\ \eqref{eq:defU} and \eqref{eq:defUtr}, 
as summarized in Table \ref{tab:U_SPT}. 
For sake of notational simplicity, below we take the convention that the ket states are included in the entries of the matrix. 
The MPS for a finite periodic system is then expressed by 
\begin{equation}\label{eq:MPS_A}
 |\Psi\rangle = {\rm Tr} \left( A^{[1]} A^{[2]}\dots A^{[L]} \right), 
\end{equation}
where $A^{[j]}$ is a ``matrix'' consisting of states on the $j$th rung in its entries. 
This ``matrix'' is related to the matrices in Eq.\ \eqref{eq:uMPS} via $A^{[j]}:=\sum_m \Lambda \Gamma_m \ket{m}_j$. 

As reviewed in Sec.\ \ref{sec:intro}, the Haldane phase appears in the simple Heisenberg ladder with $J>0$ and $\Jperp<0$. 
There, triplet states $\{ \ket{\ua\ua}, \ket{t_z}, \ket{\da\da} \}$ on the rungs collectively form a spin-$1$ Haldane state. 
An approximation to the latter is given by the exact valence bond solid ground state of the Affleck-Kennedy-Lieb-Tasaki model \cite{AKLT87,AKLT88}, 
whose MPS representation is well-known \cite{Tasaki20book,Klumper93}. 
Thus, the matrix $A$ in Eq.\ \eqref{eq:MPS_A} for the representative Haldane state on a ladder is given by 
\begin{equation}\label{eq:MPS_H}
 A_\text{H}=\frac12 \left( \ket{t_x}\sigma^x+\ket{t_y}\sigma^y+\ket{t_z}\sigma^z \right), 
\end{equation}
where the subscript ``H'' stands for ``Haldane'', $(\sigma^x,\sigma^y,\sigma^z)$ are Pauli matrices, and we use the triplet basis in Eq.\ \eqref{eq:stxyz}. 
As the Haldane* and c-Haldane* states can be viewed as the Haldane states in the basis \eqref{eq:basis_Hals} or \eqref{eq:basis_cHals}, 
we obtain the matrix $A$ for these states as 
\begin{align}
 A_\text{H*}&=\frac12 \left( i\ket{t_y}\sigma^x-i\ket{t_x}\sigma^y+\ket{s}\sigma^z \right), \label{eq:MPS_Hs}\\
 A_\text{cH*}&=\frac12 \left( -i\ket{t_y}\sigma^x+i\ket{t_x}\sigma^y+i\ket{s}\sigma^z \right). \label{eq:MPS_cHs}
\end{align}
We note that $A_\text{H}$ and $A_\text{H*}$ are real in the basis $\ket{\alpha\beta}~(\alpha,\beta=\uparrow,\downarrow)$ 
while $A_\text{cH*}$ is complex in the same basis. 
This leads to a crucial difference in the transformation under time reversal, as we shall see below. 

In terms of the ``matrix'' $A$, Eq.\ \eqref{eq:defU} for the on-site unitary transformation $\Sigma$ can be rewritten as 
\begin{equation}\label{eq:defU_A}
 \Sigma A = e^{i\theta_\Sigma} U_\Sigma^\dagger A U_\Sigma,
\end{equation}
where $\Sigma$ acts on the physical ket states in the entries of $A$ 
while $U_\Sigma$ and $U_\Sigma^\dagger$ acts on the virtual indices of the $\chi\times\chi$ matrix.  
Similarly, Eq.\ \eqref{eq:defUtr} can be rewritten as 
\begin{equation}\label{eq:defUtr_A}
 \Sigma^y A^* = e^{i\theta_{\rm TR}} U_{\rm TR}^\dagger A U_{\rm TR}. 
\end{equation}

For $A_\text{H}$, $A_\text{H*}$, and $A_\text{cH*}$ above, 
we can explicitly calculate the phase factors $e^{i\theta_\Sigma}$ and the matrices $U_\Sigma$ for different symmetries 
via Eq.\ \eqref{eq:defU_A} or \eqref{eq:defUtr_A}, as summarized in Table \ref{tab:U_SPT}. 
See Ref.\ \cite{LiuZX12} for a similar table for distinct featureless phases on a two-leg spin ladder with the $D_2\times\sigma$ and translational symmetries; 
in this reference, the Haldane and Haldane* phases are called the $t_0$ and $t_z$ phases, respectively. 
The topological indices \eqref{eq:Oxy_Oxsig_Otr} can be easily read off from Table \ref{tab:U_SPT}. 
For example, we have ${\cal O}_{\rm TR}=-1$ from the nontrivial relation $U_{\rm TR}U_{\rm TR}^*=-I$ in the Haldane and Haldane* states 
while we have ${\cal O}_{\rm TR}=+1$ from the trivial relation $U_{\rm TR}U_{\rm TR}^*=I$ in the c-Haldane* state. 


\bibliography{references}

\end{document}